\newif\ifappendixonly
\ExplSyntaxOn
\str_if_eq:eeT { \jobname } { appendix } { \appendixonlytrue }
\ExplSyntaxOff
\ifappendixonly
  \PassOptionsToClass{nonacm}{acmart}
\fi
\RequirePackage{xr}
\documentclass[acmsmall]{acmart}

\usepackage{main}
\usepackage{widows-and-orphans}
\AddToHook{cmd/thebibliography/after}{\widowpenalty=10000\relax}
\usepackage{todo}

\settoggle{arxiv}{true}

\ifdefined\conferencepaper\togglefalse{arxiv}\fi
\ifdefined\arxivsubmission\toggletrue{arxiv}\fi
\ifappendixonly
  \toggletrue{arxiv}
  \ExplSyntaxOn
  \NewDocumentCommand{\maincref}{m}
    { \exp_args:Ne \cref { paper-\clist_use:nn {#1} {,paper-} } }
  \ExplSyntaxOff
\else
  \newcommand{\maincref}[1]{\cref{#1}}
\fi
\iftoggle{arxiv}{}{%
  \IfFileExists{build/extended/main.aux}
    {\externaldocument[extended-][nocite]{build/extended/main}}
    {\PackageError{main}{Missing appendix labels. Run ./build.sh conference}
      {The script builds the extended edition first to resolve appendix citations.}}}

\AtBeginDocument{

  \theoremstyle{acmplain}
  
}

\Crefname{axiom}{Axiom}{Axioms}
\crefname{appendix}{Appendix}{Appendices}
\Crefname{appendix}{Appendix}{Appendices}

\setcopyright{cc}
\setcctype{by}
\acmDOI{10.1145/3871776}
\acmYear{2027}
\acmJournal{PACMPL}
\acmVolume{11}
\acmNumber{POPL}
\acmArticle{74}
\acmMonth{1}
\acmSubmissionID{popl27main-p702-p}
\received{2026-07-08}
\received[accepted]{2026-10-05}

\begin{document}

\title{Directed Proof-Relevant Logical Relations in Simplicial HoT{}T}
\ifappendixonly
  \subtitle{Supplementary Appendices}
\else
  \iftoggle{arxiv}{\subtitle{Extended Version with Supplementary Appendices}}{}
\begin{abstract}
Some intrinsically typed presentations of type theory use equality types in the meta-language to represent object-language
judgmental equality. In such equational syntax, one can define logical relations by assigning proof-relevant computability predicates to judgmental equivalence classes of types and terms. This approach, however, accounts for judgmental equality but not for reduction, which plays a central role in many logical-relations arguments.
This paper develops a directed version of proof-relevant logical relations in simplicial homotopy type theory, a directed type theory with an \emph{inequality} type. In simplicial type theory, we use inequality types to internalize object-language reductions and construct object syntax as a directed quotient inductive type. The central observation is that contravariant families in simplicial type theory provide exactly the proof-relevant form of closure under expansion for logical relations: computability evidence can be transported backward along reductions, with the required functoriality and universal property built in.
Using this observation, we construct a unary logical relations model with contravariant computability predicates and prove directed Boolean canonicity: every closed Boolean term reduces to either true or false. We then generalize the construction to dependent type formers and universes, where a comonadic flat modality provides the discreteness needed for type conversion and universe predicates. Finally, we extend the method to binary logical relations, separating vertical reduction from horizontal parametricity and obtaining a proof-relevant account of representation independence.
\end{abstract}
 \fi

\author{Runming Li}
\orcid{0000-0001-7600-9069}
\email{runmingl@cs.cmu.edu}

\author{Harrison Grodin}
\orcid{0000-0002-0947-3520}
\email{hgrodin@cs.cmu.edu}

\author{Robert Harper}
\orcid{0000-0002-9400-2941}
\email{rwh@cs.cmu.edu}

\affiliation{
  \institution{Carnegie Mellon University}
  \department{Computer Science Department}
  \city{Pittsburgh}
  \state{PA}
  \country{USA}
}

\ifappendixonly\else


\fi

\maketitle

\newcommand{\paperreferences}{%
  \bibliographystyle{ACM-Reference-Format}%
  \bibliography{bib/directed,bib/gluing,bib/other}%
}

\ifappendixonly\else
\section{Introduction}\label{sec:intro}

Logical relations begin with a simple idea: interpret each type by a family of
computable terms, and interpret each type former by its action on such
families. This is the pattern behind Tait's computability method
\cite{tait>1967}. A product is computable when its projections are computable;
a function is computable when it takes computable arguments to computable
results. The fundamental theorem then states that every well-typed term is
computable at its type. This method has become one of the standard tools of
programming-language semantics, used to prove properties such as normalization,
contextual equivalence, representation independence, and
noninterference.
The same idea also shapes foundational accounts of type theory itself. In the NuPRL
tradition, a computational semantics of types---closely related to
PER/logical-relations models---forms part of the basis on which the type system
is justified~\cite{constable-etal>1986,allen>thesis}.

In programming-language semantics, logical relations are often formulated
relative to a reduction relation $\rightarrow$ and its reflexive--transitive closure $\rightarrow^*$. In that setting, computation has a direction:
a term steps to, or reduces to, another term. A logical relation is then a family of
predicates defined by induction on types; for each type $A$, the predicate
$A^\bullet$ is the logical interpretation of $A$:
\[
\begin{array}{@{}l@{\;}c@{\;}l@{}}
  (-)^\bullet~(-) &:& (A : \sty) \to \stm~A \to \kw{Prop}
  \\[3pt]
  \syn{\kw{Bool}}^\bullet~(M) &\isdef&
    (M \rightarrow^* \syn{\kw{true}})
    \lor (M \rightarrow^* \syn{\kw{false}})
  \\[3pt]
  (A \mathbin{\syn{\times}} B)^\bullet~(P) &\isdef&
    A^\bullet~(\syn{\kw{fst}}~P) \land B^\bullet~(\syn{\kw{snd}}~P)
  \\[3pt]
  (A \mathbin{\syn{\Rightarrow}} B)^\bullet~(F) &\isdef&
    (M : \stm~A) \to A^\bullet~(M) \to B^\bullet~(\syn{\kw{app}}~F~M).
\end{array}
\]
The Boolean predicate says that a closed term $M$ is computable as a Boolean when it
reduces to $\syn{\kw{true}}$ or $\syn{\kw{false}}$. The inductive argument commonly
requires a closure under (head) \emph{expansion} lemma:
\[
  \text{if }
  M \rightarrow M',
  \text{ then }
  A^\bullet~(M')
  \to
  A^\bullet~(M).
\]
For Booleans, the proof composes the step $M \rightarrow M'$ with the
reduction from $M'$ to the chosen canonical Boolean. This lemma is needed to
move computability backward along computation, keeping the fundamental theorem
compatible with reduction.

A complementary route comes from category theory. Categorical gluing
\cite{mitchell-scedrov>1992,altenkirch-hofmann-streicher>1995,fiore>2002} gives a \emph{proof-relevant} account of logical
relations in which the computability predicate is not merely a proposition, but a family of \emph{types} whose inhabitants are
computability witnesses carrying non-trivial structure. This form of logical relations is widely used in modern type theory
semantics to prove
canonicity, normalization, and parametricity for a range of type theories
\cite{coquand>2018,altenkirch-kaposi>2017,kaposi-huber-sattler>2019,bocquet-kaposi-sattler>2023}.
In this approach, the syntactic component usually consists of judgmental
equivalence classes of types and terms, rather than raw terms equipped with a
transition system. For example, the product fragment contains constructors and
equations of the following shape:
\[
\begin{array}{@{}l@{\;}c@{\;}l@{}}
  \syn{\kw{pair}} &:& \stm~A \to \stm~B \to \stm~(A \mathbin{\syn{\times}} B) \\
  \syn{\kw{fst}} &:& \stm~(A \mathbin{\syn{\times}} B) \to \stm~A \\
  \syn{\kw{snd}} &:& \stm~(A \mathbin{\syn{\times}} B) \to \stm~B
\end{array}
\qquad
\begin{array}{@{}l@{\;}c@{\;}l@{}}
  \syn{\times_{\beta_1}} &:&
    \syn{\kw{fst}}~(\syn{\kw{pair}}~M~N) = M \\
  \syn{\times_{\beta_2}} &:&
    \syn{\kw{snd}}~(\syn{\kw{pair}}~M~N) = N \\
  \syn{\times_{\eta}} &:&
    \syn{\kw{pair}}~(\syn{\kw{fst}}~P)~(\syn{\kw{snd}}~P) = P .
\end{array}
\]
The logical relation assigns a proof-relevant predicate to each type.
For example, the product predicate is defined by
\[
\begin{array}{@{}l@{\;}c@{\;}l@{}}
  (-)^\bullet~(-) &:& (A : \sty) \to \stm~A \to \Univ
  \\[3pt]
  (A \mathbin{\syn{\times}} B)^\bullet~(P)
    &\isdef&
    A^\bullet~(\syn{\kw{fst}}~P)
    \times
    B^\bullet~(\syn{\kw{snd}}~P).
\end{array}
\]
For a term $M : \stm~A$, we write $M^\bullet$ for its computability witness of type $A^\bullet~(M)$. For example,
the computability evidence for a pair has the following type:
\[
\begin{array}{@{}l@{\;}c@{\;}l@{}}
  (-)^\bullet &:& (M : \stm~A) \to A^\bullet~(M) \\[3pt]
  (\syn{\kw{pair}}~M~N)^\bullet
    &:&
    (A \mathbin{\syn{\times}} B)^\bullet~(\syn{\kw{pair}}~M~N)
  \\[3pt]
    &=&
    A^\bullet~(\syn{\kw{fst}}~(\syn{\kw{pair}}~M~N))
    \times
    B^\bullet~(\syn{\kw{snd}}~(\syn{\kw{pair}}~M~N)).
\end{array}
\]
Given $M : \stm~A$ and $N : \stm~B$, the induction hypotheses provide
$M^\bullet : A^\bullet~M$ and
$N^\bullet : B^\bullet~N$ as computability evidence for the components.
The product $\beta$-equations identify $\syn{\kw{fst}}$ and $\syn{\kw{snd}}$ of
$\syn{\kw{pair}}~M~N$ with $M$ and $N$, respectively. Thus the computability evidence for a
pair is the pair of computability witnesses, modulo the two $\beta$-equations:
\[
  (\syn{\kw{pair}}~M~N)^\bullet
  \isdef
  (M^\bullet, N^\bullet).
\]
The projection evidence is obtained by extracting the relevant component:
\[
  (\syn{\kw{fst}}~P)^\bullet \isdef \pi_1(P^\bullet)
  \qquad
  (\syn{\kw{snd}}~P)^\bullet \isdef \pi_2(P^\bullet) .
\]
In particular, computability evidence respects equations. For example, in the case of $\syn{\times_{\beta_1}}$, the
computability evidence for $\syn{\kw{fst}}~(\syn{\kw{pair}}~M~N)$ is equal to the computability evidence for $M$:
\[
  (\syn{\kw{fst}}~(\syn{\kw{pair}}~M~N))^\bullet
  = \pi_1((\syn{\kw{pair}}~M~N)^\bullet)
  = \pi_1(M^\bullet, N^\bullet)
  = M^\bullet .
\]

Proof relevance matters here for two reasons. First, the computability of a universe should contain the computability predicate for each type in that universe. That extra structure is unavailable
when computability predicates are proof-irrelevant. Second, the treatment of equations, such as the $\beta$-law above, requires a comparison of computability evidence.
Gluing packages these obligations into one algebraic construction, and the fundamental
theorem becomes the construction of a model.

The two perspectives above emphasize different aspects of logical relations. Operational logical relations treat
computation as directed structure: terms reduce, and computability must be closed backward along those reductions.
Gluing, on the other hand, treats computability evidence as part of the semantics: equations must compare
terms and the evidence. This paper asks whether these two aspects can coexist. The guiding question of this paper is therefore:
\[
  \begin{gathered}
  \textit{Can proof-relevant gluing for logical relations account for directed reductions?}
  \end{gathered}
\]

\subsection{From Equalities to Inequalities}

The equational approach represents object-language judgmental equalities by
meta-level equalities, in effect presenting the syntax as a quotient by those
equations. This is convenient because equality in the meta-language already has
the structural behavior expected of judgmental equality: reflexivity,
transitivity, symmetry, and congruence. Compatibility with type and term formers
is therefore inherited from meta-level congruence. For products, the following left-hand
compatibility rule is recovered from the right-hand meta-level congruence term:
\[
  \begin{array}{@{}c@{\hspace{3em}}l@{}}
    \inferrule{
      \Gamma \vdash M \equiv M' : A \\
      \Gamma \vdash N \equiv N' : B \\
    }{
      \Gamma \vdash \syn{\kw{pair}}~M~N \equiv \syn{\kw{pair}}~M'~N' : A \mathbin{\syn{\times}} B
    }
    &
    \begin{array}[c]{@{}l@{\;}c@{\;}l@{}}
      \kw{cong}_2~\syn{\kw{pair}} & : &
        M = M' \to N = N' \to \\
      & & \syn{\kw{pair}}~M~N = \syn{\kw{pair}}~M'~N'
    \end{array}
  \end{array}
\]
The exact use of these equations depends on the chosen meta-language. In extensional
type theory, equality reflection turns equality proofs into judgmental
equalities. In intensional type theory and homotopy type theory, identity types
or path types instead provide propositional equations, and constructions must
transport along those equations.

Unlike judgmental equality, reduction is fundamentally directed. If representing judgmental equality calls for equality types in the meta-language, then representing reduction internally calls for an asymmetric analogue of equality. Directed type
theories~\cite{licata-harper>2011} typically provide such structure in the form
of homomorphism or inequality types. Among variants of directed type theories, this paper works in
the simplicial homotopy type theory of \citet{riehl-shulman>2017}, where inequality types $x \leq_X x'$ serve as
the directed counterpart of equality types. The technical details are recalled
in \cref{sec:simplicial-background}. For now, the important point is that inequalities are reflexive and
preserved by functions, as equalities are, but need not be symmetric.

In the equational case, the syntax is constructed intrinsically via a quotient inductive-inductive presentation
of type theory~\cite{altenkirch-kaposi>2016,altenkirch-capriotti-dijkstra-kraus-nordvall-forsberg>2018,kaposi-kovacs-altenkirch>2019}.
The present paper generalizes this idea by also adding directed constructors, which generate inequalities rather than equalities, to those quotients.
Thus, instead of imposing a product computation equation such as
$\syn{\kw{fst}}~(\syn{\kw{pair}}~M~N) = M$, we add directed constructors
that internalize the reductions:
\[
\begin{array}{@{}l@{\;}c@{\;}l@{\qquad\qquad}l@{\;}c@{\;}l@{}}
  \syn{\times_{\beta_1}} & : &
    \syn{\kw{fst}}~(\syn{\kw{pair}}~M~N) \leq M
  &
  \syn{\times_{\beta_2}} & : &
    \syn{\kw{snd}}~(\syn{\kw{pair}}~M~N) \leq N .
\end{array}
\]
With this syntax in place, the goal is a canonicity result by logical
relations: every closed Boolean term $M$ reduces to a canonical Boolean,
$M \leq \syn{\kw{true}}$ or $M \leq \syn{\kw{false}}$.

\subsection{Contravariance as Proof-Relevant Expansion}\label{sec:intro:reductions}

Returning to the discussion of expansion: a computability predicate $A^\bullet$ should be able to transport
evidence backward along reductions: each $r : M \leq M'$ should induce a map
$r^* : A^\bullet~(M') \to A^\bullet~(M)$.
For a proof-irrelevant logical relation, such a map is sufficient.
However, proof relevance imposes coherence conditions on the map
$r^*$, requiring it to be
\emph{universal} in an appropriate sense.

The key observation of this work is that simplicial type theory already provides
this coherent form of expansion. Its notion of a \emph{contravariant family} was
developed to study synthetic fibrations, but it has exactly the structure needed
for proof-relevant logical relations. At a high level, a contravariant family is a type family equipped with backward transport along
inequalities. In short:
\[
  \begin{gathered}
    \textit{Contravariance is proof-relevant closure under expansion.}
  \end{gathered}
\]
The rest of the paper develops this observation into a directed version of
proof-relevant logical relations. Each type is equipped with a contravariant
computability predicate, and the resulting fundamental theorem transports
computability evidence backward along reductions.

\subsection{Simplicial Homotopy Type Theory}\label{sec:intro:dtt}
This work takes place in simplicial homotopy type theory~\cite{riehl-shulman>2017},
a directed extension of homotopy type theory (HoTT)~\cite{hott>2013}. We begin by
recalling a few notions in HoTT that will be used later. In HoTT, path types $x =_X x'$ express proof-relevant,
higher-dimensional, typal equality: between two points, there may
be many distinct paths. When the type $X$ is clear from context, we omit the
subscript and write $x = x'$.

\subsubsection{Homotopy Levels}
This higher path structure organizes types by homotopy level. A type is
\emph{contractible} when it has a distinguished point, called its center, to which
every other point is equal; a type is a \emph{proposition} when any two of its points
are equal; and a type is a \emph{set} when its path types are propositions, \ie when any two
paths between the same two points are equal. We use the following definitions,
and write $\kw{center}(c)$ for the first projection of $c : \kw{isContr}~X$.
\[
\begin{array}{@{}c@{\hspace{3em}}c@{}}
  \begin{array}[t]{@{}l@{\;}c@{\;}l@{}}
    \kw{isContr}~X
    &\isdef&
    \displaystyle\sum_{x : X} (x' : X) \to x =_X x'
  \end{array}
  &
  \begin{array}[t]{@{}l@{\;}c@{\;}l@{}}
    \kw{isProp}~X
    &\isdef&
    (x~x' : X) \to x =_X x'
    \\[3pt]
    \kw{isSet}~X
    &\isdef&
    (x~x' : X) \to \kw{isProp}~(x =_X x').
  \end{array}
\end{array}
\]

\subsubsection{Equivalences}
Equivalences between types can be expressed through contractible fibers. For a
map $f : X \to Y$, the fiber over $y : Y$ records the preimages of $y$ under
$f$:
\[
  \kw{fib}_f(y) \isdef \sum_{x : X} f~x =_Y y,
  \qquad
  \kw{isEquiv}~f \isdef (y : Y) \to \kw{isContr}~(\kw{fib}_f(y)).
\]
Thus a map is an equivalence when every element of the codomain has a unique
preimage, up to paths. This definition packages both inverse data and the
coherence laws relating that inverse to the original map. The notation $X\simeq Y$ denotes the type of equivalences from $X$ to $Y$.

\subsubsection{Higher Inductive Types}
Homotopy type theory also provides higher inductive types (HITs). Unlike
ordinary inductive types, a HIT may include both point constructors and path
constructors.
Path constructors impose the quotient equations, making HITs a natural way to present
syntax modulo equations~\cite{altenkirch-kaposi>2016}.

\subsubsection{Simplicial HoTT}
Simplicial homotopy type theory extends HoTT with a directed interval. Maps out
of this interval behave as directed paths, and we write the type of such paths
from $x$ to $x'$ in $X$ as inequalities $x \leq_X x'$. In the simplicial type theory
literature, this is commonly called the hom type $\kw{hom}_X(x,x')$,
whose elements are synthetic morphisms~\cite{riehl-shulman>2017,gratzer-weinberger-buchholtz>2024,gratzer-weinberger-buchholtz>2024-yoneda,gratzer-weinberger-buchholtz>2026}.
In this work, we use inequality notation to emphasize its role as an asymmetric
analogue of equality types in dependent type theory. These inequalities are
asymmetric, because the directed interval has no reversal operation, but every
function is guaranteed to be \emph{monotone}, the directed analogue of congruence.
The necessary definitions and theorems from simplicial type theory
are recalled as needed throughout the paper; for a complete treatment, see
\citet{riehl-shulman>2017}.
The extension developed by \citet{gratzer-weinberger-buchholtz>2024} adds
modalities from \citet{gratzer>thesis} to simplicial type theory. In particular,
it includes the flat modality $\flat$ from crisp type
theory~\cite{shulman>2018-brouwer,licata-orton-pitts-spitters>2018}, used in
\cref{sec:universe}.

\subsection{Contributions and Synopsis}\label{sec:intro:idea}

The paper develops directed proof-relevant logical relations through the
following contributions.

\begin{enumerate}[leftmargin=*]
\item \textbf{Directed quotient syntax.}
  \Cref{sec:syntax} presents the syntax as a directed quotient
  inductive-inductive type, where reductions are modeled by directed
  inequalities.
\item \textbf{Contravariance as proof-relevant expansion.}
  \Cref{sec:gluing} identifies contravariance as the proof-relevant form of
  closure under expansion, requiring each computability predicate to carry a
  coherent contravariant structure.
\item \textbf{Directed canonicity by gluing.}
  The unary logical relation of \cref{sec:gluing} proves directed Boolean
  canonicity: every closed Boolean term $M$ satisfies either
  $M \leq \syn{\kw{true}}$ or $M \leq \syn{\kw{false}}$.
\item \textbf{Mechanization.}
  As a proof of concept, \cref{sec:syntax,sec:gluing} are mechanized in
  Cubical Agda, including the directed syntax, the contravariance condition and
  its properties, and the logical relations model with products, functions, and
  Booleans. We mark definitions, lemmas, and
  constructions covered by the mechanization with \AgdaFormalized{} throughout the paper.
\item \textbf{Universes and dependency.}
  \Cref{sec:gluing} treats non-dependent type formers first, while
  \cref{sec:universe} adds dependent type formers and universes. The universe
  predicate requires the flat modality $\flat$ to establish contravariance.
\item \textbf{Binary parametricity.}
  \Cref{sec:binary} adapts the construction to binary logical relations and
  parametricity, separating vertical reductions from horizontal parametricity
  witnesses and illustrating the result with a queue example.
\end{enumerate}
\section{Modeling Reduction in Syntax}\label{sec:syntax}

Typically, the syntax of a type theory can be described as a signature in some
logical framework. The judgmental structure and the type and term formers are
constants in the signature, while judgmental equalities are represented using
the equality notion supplied by the meta-language. One such presentation is the
structure of a category with families (CwF)~\cite{dybjer>1996}.

The CwF part of the signature supplies the ambient judgments: contexts,
substitutions, types in a context, and terms of a type in a context.
Object-language syntax is set in \syn{red} throughout.
\[
\begin{array}{@{}c@{\hspace{3em}}c@{}}
  \begin{array}[t]{@{}l@{\;}c@{\;}l@{}}
    \sctx & : & \Univ
    \\
    \ssub & : & \sctx \to \sctx \to \Univ
  \end{array}
  &
  \begin{array}[t]{@{}l@{\;}c@{\;}l@{}}
    \sty & : & \sctx \to \Univ
    \\
    \stm & : & (\Gamma : \sctx) \to
      \sty~\Gamma \to \Univ
  \end{array}
\end{array}
\]
Here $\sctx$ is the type of contexts,
$\ssub~\Delta~\Gamma$ is the type of substitutions assigning terms
in context $\Delta$ to variables of context $\Gamma$, $\sty~\Gamma$ is the type of types in
context $\Gamma$, and $\stm~\Gamma~A$ is the type of terms of type $A$
in context $\Gamma$.
Product types, for example, are then added as an extension of this CwF signature. The left
half below lists the type and term formers; the right half lists the
judgmental equations.
\[
\begin{array}{@{}c@{\hspace{3em}}c@{}}
  \begin{array}[t]{@{}l@{\;}c@{\;}l@{}}
    -\mathbin{\syn{\times}}- & : &
      \sty~\Gamma \to
      \sty~\Gamma \to \sty~\Gamma
    \\
    \syn{\kw{pair}} & : &
      \stm~\Gamma~A \to
      \stm~\Gamma~B \to
      \stm~\Gamma~(A \mathbin{\syn{\times}} B)
    \\
    \syn{\kw{fst}} & : &
      \stm~\Gamma~(A \mathbin{\syn{\times}} B) \to
      \stm~\Gamma~A
    \\
    \syn{\kw{snd}} & : &
      \stm~\Gamma~(A \mathbin{\syn{\times}} B) \to
      \stm~\Gamma~B
  \end{array}
  &
  \begin{array}[t]{@{}l@{\;}c@{\;}l@{}}
    \syn{\times_{\beta_1}} & : &
      \syn{\kw{fst}}~(\syn{\kw{pair}}~M~N) = M
    \\
    \syn{\times_{\beta_2}} & : &
      \syn{\kw{snd}}~(\syn{\kw{pair}}~M~N) = N
    \\
    \syn{\times_{\eta}} & : &
      \syn{\kw{pair}}~(\syn{\kw{fst}}~P)~(\syn{\kw{snd}}~P) = P
  \end{array}
\end{array}
\]

From here on, the aim is to model judgmental reduction. To do that, the equations above are replaced by directed constructors, represented internally by inequality types.

\subsection{Background on Simplicial Type Theory}\label{sec:simplicial-background}

We first recall the basic structure of simplicial type theory, which provides these types.

\subsubsection{The Directed Interval and Inequality Types}

Simplicial type theory extends HoTT~\cite{hott>2013} with one
primitive type: a directed interval $\mathbbm{2}$ with two endpoints
$\mathsf{i0}$ and $\mathsf{i1}$. Its direction is part of the structure:
$\mathbbm{2}$ is a bounded total order with $\mathsf{i0}$ as minimum and $\mathsf{i1}$ as maximum.
A map out of $\mathbbm{2}$ is a directed path in the target type.

\begin{definition}[Inequality type \AgdaFormalized]\label{def:hom}
For $x, x' : X$, let the type of inequalities%
\footnote{\citet{riehl-shulman>2017} define inequality types using
extension types, so that the endpoint conditions such as $p~\mathsf{i0} = x$
are treated judgmentally. For this paper, it is enough to use the resulting
inequality types and their expected structural principles, so we do not develop
the machinery of extension types.}
between $x$ and $x'$ be
\[
  x \leq_X x' \isdef \sum_{p : \mathbbm{2} \to X} (p~\mathsf{i0} = x) \times (p~\mathsf{i1} = x').
\]
\end{definition}
For every $x : X$, there is a reflexivity term $\kw{id}_x : x \leq_X x$ given by the constant path at $x$.
Unlike identity types, inequality types are not symmetric. They nevertheless have
the structural behavior needed to play the role of judgmental congruence. In
particular, every function $f : X \to Y$ is automatically monotone: it acts on
inequalities functorially.
\[
  \kw{mono}_f : (x \leq_X x') \to (f~x \leq_Y f~x')
\]
We sometimes simply write $f$ for $\kw{mono}_f$ when the functorial action is clear.

\subsubsection{The Reflective Subuniverse of Preorders}\label{sec:reflective-subuniverse}

Inequality types in simplicial type theory model morphisms in $\infty$-categories. In this work, we focus on preorders, a particular kind of category. There are three standard properties of preorders that need not hold for an arbitrary type $X$:
\begin{enumerate}
  \item inequality should be transitive (given $p : x \leq_X x'$ and $q : x' \leq_X x''$, there is $p \cdot q : x \leq_X x''$);
  \item inequality should be a proposition (for all $p,q : x \leq_X x'$, it is the case that $p =_{x \leq_X x'} q$); and
  \item equality should be a proposition (for all $p,q : x =_X x'$, it is the case that $p =_{x =_X x'} q$).
\end{enumerate}
However, there is a reflective subuniverse $\mathcal{U}_p \hookrightarrow \Univ$ of \emph{preorders} consisting of exactly the types $X$ for which these properties hold.
In this work, we make use of this $\mathcal{U}_p$, combining the reflective subuniverse of Segal types \citep{riehl-shulman>2017,rijke-shulman-spitters>2020} and a restriction to propositional inequality \citep{grodin-niu-sterling-harper>2024}.
The full construction is recalled in \citeappendix{appendix:preorders}.

\subsubsection{Inequality Principles}

Inequality propagates through type structures naturally; for example, inequality at dependent function types satisfies a directed analogue of function extensionality.
\begin{lemma}[Directed function extensionality; {\protect\citealp[Prop.~6.3]{riehl-shulman>2017}} \AgdaFormalized]\label{lem:fun-ext}
  For $f,g : (x : X) \to Y(x)$, the canonical map $(f \leq g) \to ((x : X) \to (f~x \leq_{Y(x)} g~x))$
  is an equivalence.
\end{lemma}

Similarly, inequality at dependent sums satisfies a directed analogue of the structure identity principle:
an inequality at a $\Sigma$-type is an inequality in the base and a dependent inequality over it.

\begin{definition}[Dependent inequality type \AgdaFormalized]\label{def:dependent-hom}
For $p : x \leq_X x'$, $y : Y(x)$, and $y' : Y(x')$, the type of dependent inequalities from $y$ to $y'$ lying over
$p$ is defined as\footnote{Here $p : x \leq_X x'$ is implicitly coerced to a map
${p : \mathbbm{2} \to X}$; the transports induced by $p$ in
$y \leq_{Y(p)} y'$ are also left implicit.}
\[
  y \leq_{Y(p)} y'
  \isdef
  \sum_{q : (i : \mathbbm{2}) \to Y(p~i)}
    (q~\mathsf{i0} = y) \times (q~\mathsf{i1} = y').
\]
This is similar to
$\kw{PathP}$ in cubical type
theories~\cite{cohen-coquand-huber-mortberg>2018,angiuli-brunerie-coquand-harper-favonia-licata>2021}.
\end{definition}

\begin{lemma}[Inequalities at $\Sigma$-types \AgdaFormalized]\label{lem:sigma-hom}
Given $Y : X \to \Univ$, $x,x' : X$, $y : Y(x)$, and $y' : Y(x')$, there is an equivalence
\[
  (x,y) \leq_{\sum_{x:X} Y(x)} (x',y')
  \simeq
  \Bigl(\sum_{p : x \leq_X x'} y \leq_{Y(p)} y'\Bigr).
\]
\end{lemma}

\subsection{Judgmental Reduction as Directed Structure \AgdaFormalized}\label{sec:judgmental-reduction}

To model judgmental reduction, we replace the equalities in the ordinary
signature by inequalities. For products, this turns the $\beta$-rules into directed constructors. The left column writes these rules
in operational notation; the right column gives their internal form as elements
of inequality types:
\[
  \begin{array}{@{}c@{\hspace{3em}}l@{\;}c@{\;}l@{}}
      \syn{\kw{fst}}~(\syn{\kw{pair}}~M~N) \rightarrow_{\beta} M
    &
    \syn{\times_{\beta_1}} & : & \syn{\kw{fst}}~(\syn{\kw{pair}}~M~N) \leq M
    \\
      \syn{\kw{snd}}~(\syn{\kw{pair}}~M~N) \rightarrow_{\beta} N
    &
    \syn{\times_{\beta_2}} & : & \syn{\kw{snd}}~(\syn{\kw{pair}}~M~N) \leq N
  \end{array}
\]
Congruence for reductions is then inherited from monotonicity. For instance,
the congruence rule for $\syn{\kw{fst}}$ is represented internally by applying
$\kw{mono}$ to the projection function:
\[
  \begin{array}{@{}c@{\hspace{3em}}l@{\;}c@{\;}l@{}}
    \inferrule{
        P \rightarrow_{\beta} P'
    }{
        \syn{\kw{fst}}~P \rightarrow_{\beta} \syn{\kw{fst}}~P'
    }
    &
    \kw{mono}_{\syn{\kw{fst}}} & : &
      P \leq P' \to \syn{\kw{fst}}~P \leq \syn{\kw{fst}}~P'
  \end{array}
\]
To ensure that such inequalities of terms (and substitutions) are well-behaved, we require that the families of substitutions and terms be valued in the universe of preorders.
\[
\begin{array}{@{}c@{\hspace{3em}}c@{}}
  \begin{array}[t]{@{}l@{\;}c@{\;}l@{}}
    \ssub & : & \sctx \to \sctx \to \mathcal{U}_p
  \end{array}
  &
  \begin{array}[t]{@{}l@{\;}c@{\;}l@{}}
    \stm & : & (\Gamma : \sctx) \to
      \sty~\Gamma \to \mathcal{U}_p
  \end{array}
\end{array}
\]
For now, we leave $\sctx$ and $\sty$ unaltered; we will restrict their target universes in \cref{sec:universe} to accommodate a universe of types.

Apart from these changes, the syntax retains the usual CwF structure; see, for example,
\citet{kaposi-huber-sattler>2019}. In particular, the equations of the
substitution calculus, together with the naturality equations for type and term
formers, are still represented by equality in the metatheory, not by
inequalities. The CwF substitution structure and non-dependent product
types are presented in \cref{fig:directed-cwf-product}. We omit the standard CwF substitution equations such
as $\ssubst{A}{\tau \scomp \sigma} = \ssubst{\ssubst{A}{\tau}}{\sigma}$ for brevity. The boxed inequalities highlight
the directed product reductions, and the
remaining laws are equalities. In HoTT, these equations are paths rather than definitional equalities, so
applying them requires transport along those paths, which we leave implicit
throughout the paper.

\begin{figure}[tbp]
\centering
\(\declkw{record}~\kw{DirectedCwF}~\declkw{where}\)\par
\begin{minipage}[t]{0.53\textwidth}
\centering
\(\displaystyle
  \begin{array}[t]{@{}l@{\;}c@{\;}l@{}}
    \sctx & : & \Univ \\
    \sty & : & \sctx \to \Univ \\
    \ssub & : & \sctx \to \sctx \to \ineqbox{\mathcal{U}_p} \\
    \stm & : & (\Gamma : \sctx) \to \sty~\Gamma \to \ineqbox{\mathcal{U}_p} \\
    \sidt & : & \ssub~\Gamma~\Gamma \\
    -\scomp- & : &
      \begin{array}[t]{@{}l@{}}
        \ssub~\Theta~\Delta \to \ssub~\Gamma~\Theta \to
        \ssub~\Gamma~\Delta
      \end{array} \\
    \syn{\emp} & : & \sctx \\
    \seps_\Gamma & : & \ssub~\Gamma~\syn{\emp} \\
    -\sext- & : &
      (\Gamma : \sctx) \to \sty~\Gamma \to \sctx \\
    \spp & : & \ssub~(\Gamma \sext A)~\Gamma \\
    \sqq & : & \stm~(\Gamma \sext A)~(\ssubst{A}{\spp}) \\
    \ssubst{-}{-} & : &
      \sty~\Delta \to \ssub~\Gamma~\Delta \to \sty~\Gamma \\
    \ssubst{-}{-} & : &
      \begin{array}[t]{@{}l@{}}
        \stm~\Delta~A \to
        (\sigma : \ssub~\Gamma~\Delta) \to
        \stm~\Gamma~(\ssubst{A}{\sigma})
      \end{array} \\
    \spair{-}{-} & : &
      \begin{array}[t]{@{}l@{}}
        (\sigma : \ssub~\Gamma~\Delta) \to
        \stm~\Gamma~(\ssubst{A}{\sigma}) \to \\
        \ssub~\Gamma~(\Delta \sext A)
      \end{array}
  \end{array}
\)
\end{minipage}\hfill%
\begin{minipage}[t]{0.47\textwidth}
\centering
\(\displaystyle
  \begin{array}[t]{@{}l@{\;}c@{\;}l@{}}
    -\mathbin{\syn{\times}}- & : &
      \sty~\Gamma \to \sty~\Gamma \to \sty~\Gamma \\
    \syn{\kw{pair}} & : &
      \stm~\Gamma~A \to
      \stm~\Gamma~B \to
      \stm~\Gamma~(A \mathbin{\syn{\times}} B) \\
    \syn{\kw{fst}} & : &
      \stm~\Gamma~(A \mathbin{\syn{\times}} B) \to \stm~\Gamma~A \\
    \syn{\kw{snd}} & : &
      \stm~\Gamma~(A \mathbin{\syn{\times}} B) \to \stm~\Gamma~B
    \\[3pt]
    \syn{\times_{[]}} & : &
      \ssubst{(A \mathbin{\syn{\times}} B)}{\sigma}
      = \ssubst{A}{\sigma} \mathbin{\syn{\times}} \ssubst{B}{\sigma} \\
    \syn{\kw{pair}_{[]}} & : &
      \ssubst{(\syn{\kw{pair}}~M~N)}{\sigma}
      = \syn{\kw{pair}}~(\ssubst{M}{\sigma})~(\ssubst{N}{\sigma}) \\
    \syn{\kw{fst}_{[]}} & : &
      \ssubst{(\syn{\kw{fst}}~P)}{\sigma}
      = \syn{\kw{fst}}~(\ssubst{P}{\sigma}) \\
    \syn{\kw{snd}_{[]}} & : &
      \ssubst{(\syn{\kw{snd}}~P)}{\sigma}
      = \syn{\kw{snd}}~(\ssubst{P}{\sigma}) \\[3pt]
    \syn{\times_{\beta_1}} & : &
      \ineqbox{\syn{\kw{fst}}~(\syn{\kw{pair}}~M~N) \leq M} \\
    \syn{\times_{\beta_2}} & : &
      \ineqbox{\syn{\kw{snd}}~(\syn{\kw{pair}}~M~N) \leq N}
  \end{array}
\)
\end{minipage}
\caption{Directed CwF operations (left; $\syn{\emp}$ is the empty context, $\Gamma \sext A$ extends $\Gamma$ by a variable of type $A$, $\spp$ weakens, and $\sqq$ is the rightmost variable) and product types and reductions (right). Boxes mark the changes from the standard equational CwF: $\ssub$ and $\stm$ land in preorders, and reductions replace equations.}
\Description{The left column lists the fields of a directed CwF record. The right column lists product constructors, substitution equations, and directed beta reductions.}
\label{fig:directed-cwf-product}
\vspace{4pt}\figrule
\end{figure}

\subsection{Initial Directed CwF via Directed Quotient Inductive-Inductive Types}\label{sec:syntax:initial-directed-cwf}

Elements of the CwF signature are models of the type theory. In particular, the
initial model gives the inductively generated syntax. To construct this syntax,
the ordinary quotient inductive-inductive type (QIIT) presentation of type theory~\cite{altenkirch-kaposi>2016,
altenkirch-capriotti-dijkstra-kraus-nordvall-forsberg>2018} is generalized to the directed setting.
In HoTT, higher inductive types (HITs) generate not only points, but also paths
between those points. A set-truncated HIT
is a quotient inductive type (QIT). This is the mechanism used in the QIIT
presentation of type theory to internalize judgmental equalities as paths in the
syntax~\cite{altenkirch-kaposi>2016}, where terms are quotiented by judgmental
equalities. To internalize judgmental
reductions, the inductive definition must also allow directed
constructors. We call the resulting notion a \emph{directed quotient
inductive-inductive type} (directed QIIT).

A schematic directed QIIT fragment for the product syntax is shown in
\cref{fig:product-qiit} as four mutually defined higher inductive types.
\begin{figure}[tbp]
\[
\makebox[\textwidth][c]{$\begin{array}{@{}l@{\hspace{1em}}l@{}}
\begin{array}[t]{@{\guiderule{oc-red-9}}l@{\;}c@{\;}l@{}}
\multicolumn{3}{@{\guiderule{oc-red-9}}l@{}}{
  \declkw{Inductive}~\sctx : \Univ~\declkw{where}} \\
\quad \syn{\emp} & : & \sctx \\
\quad -\sext- & : &
  (\Gamma : \sctx) \to \sty~\Gamma \to \sctx
\\ \noalign{\vskip 8pt}
\multicolumn{3}{@{\guiderule{oc-red-9}}l@{}}{
  \declkw{Inductive}~\ssub :
  \sctx \to \sctx \to \ineqbox{\mathcal{U}_p}~\declkw{where}} \\
\quad \sidt & : & \ssub~\Gamma~\Gamma \\
\quad -\scomp- & : &
  \begin{array}[t]{@{}l@{}}
    \ssub~\Theta~\Delta \to
    \ssub~\Gamma~\Theta \to \\
    \ssub~\Gamma~\Delta
  \end{array} \\
\quad \cdots
\\ \noalign{\vskip 8pt}
\multicolumn{3}{@{\guiderule{oc-red-9}}l@{}}{
  \declkw{Inductive}~\sty :
  \sctx \to \Univ~\declkw{where}} \\
\quad \ssubst{-}{-} & : &
  \sty~\Delta \to \ssub~\Gamma~\Delta \to \sty~\Gamma \\
\quad \cdots \\
\quad -\mathbin{\syn{\times}}- & : &
  \sty~\Gamma \to \sty~\Gamma \to \sty~\Gamma \\
\quad \syn{\times_{[]}} & : &
  \begin{array}[t]{@{}l@{}}
    \ssubst{(A \mathbin{\syn{\times}} B)}{\sigma}
    = \ssubst{A}{\sigma} \mathbin{\syn{\times}} \ssubst{B}{\sigma}
  \end{array} \\
\quad \cdots
\end{array}
&
\begin{array}[t]{@{\guiderule{oc-red-9}}l@{\;}c@{\;}l@{}}
\multicolumn{3}{@{\guiderule{oc-red-9}}l@{}}{
  \declkw{Inductive}~\stm :
  (\Gamma : \sctx) \to \sty~\Gamma \to
  \ineqbox{\mathcal{U}_p}~\declkw{where}} \\
\quad \ssubst{-}{-} & : &
  \begin{array}[t]{@{}l@{}}
    \stm~\Delta~A \to
    (\sigma : \ssub~\Gamma~\Delta) \to \\
    \stm~\Gamma~(\ssubst{A}{\sigma})
  \end{array} \\
\quad \cdots \\
\quad \syn{\kw{pair}} & : &
  \begin{array}[t]{@{}l@{}}
    \stm~\Gamma~A \to \stm~\Gamma~B \to
    \stm~\Gamma~(A \mathbin{\syn{\times}} B)
  \end{array} \\
\quad \syn{\kw{fst}} & : &
  \stm~\Gamma~(A \mathbin{\syn{\times}} B) \to \stm~\Gamma~A \\
\quad \syn{\kw{snd}} & : &
  \stm~\Gamma~(A \mathbin{\syn{\times}} B) \to \stm~\Gamma~B \\
\quad \cdots \\
\quad \syn{\kw{pair}_{[]}} & : &
  \begin{array}[t]{@{}l@{}}
    \ssubst{(\syn{\kw{pair}}~M~N)}{\sigma} =
    \syn{\kw{pair}}~(\ssubst{M}{\sigma})~(\ssubst{N}{\sigma})
  \end{array} \\
\quad \syn{\kw{fst}_{[]}} & : &
  \begin{array}[t]{@{}l@{}}
    \ssubst{(\syn{\kw{fst}}~P)}{\sigma} =
    \syn{\kw{fst}}~(\ssubst{P}{\sigma})
  \end{array} \\
\quad \syn{\kw{snd}_{[]}} & : &
  \begin{array}[t]{@{}l@{}}
    \ssubst{(\syn{\kw{snd}}~P)}{\sigma} =
    \syn{\kw{snd}}~(\ssubst{P}{\sigma})
  \end{array} \\
\quad \cdots \\
\quad \syn{\times_{\beta_1}} & : &
  \ineqbox{\syn{\kw{fst}}~(\syn{\kw{pair}}~M~N) \leq M} \\
\quad \syn{\times_{\beta_2}} & : &
  \ineqbox{\syn{\kw{snd}}~(\syn{\kw{pair}}~M~N) \leq N} \\
\quad \cdots
\end{array}
\end{array}$}
\]
\caption{A schematic directed QIIT for product syntax.}
\Description{Four mutually defined inductive types for contexts, substitutions, types, and terms, including product constructors, substitution equations, and directed beta reductions.}
\label{fig:product-qiit}
\vspace{4pt}\figrule
\end{figure}

\subsubsection{Directed Higher Inductive Types \AgdaFormalized}\label{sec:syntax:directed-hit}

The resulting inductive definition is a mixture of point constructors, such as
  $\syn{\kw{pair}}$ and $\syn{\kw{fst}}$, equality constructors, such as
  $\syn{\times_{[]}}$ and $\syn{\kw{fst}_{[]}}$, and inequality constructors,
  such as $\syn{\times_{\beta_1}}$ and $\syn{\times_{\beta_2}}$. The first two
  kinds are standard in HITs, but
  directed constructors are less so.
In directed type
theories, \citet[Section~4.5.1]{grodin-niu-sterling-harper>2024} and \citet{weaver-licata>2020} already
introduce some instances of directed HITs. The role of the present section is to explain how to use the same directed infrastructure for the initial syntax of the object theory.

The key point is that a directed constructor is syntactic sugar for ordinary
higher constructors involving the directed interval. By \cref{def:hom}, a directed constructor $\syn{c} : x \leq_X x'$ expands into a point constructor $\syn{c_{\kw{rel}}}$ and two equality constructors $\syn{c_{\kw{left}}}$ and $\syn{c_{\kw{right}}}$, all of which are accepted in typical higher inductive type syntax, as follows:
\[
\begin{array}{@{}l@{\qquad}c@{\qquad}l@{}}
\begin{array}[t]{@{\guiderule{oc-red-9}}l@{}}
  \declkw{Inductive}~X : \mathcal{U}_p~\declkw{where} \\
  \quad \cdots \\
  \quad \syn{c} : x \leq_X x'
\end{array}
&
\begin{array}[t]{@{}c@{}} \\ \\ \leadsto \end{array}
&
\begin{array}[t]{@{\guiderule{oc-red-9}}l@{\;}c@{\;}l@{}}
\multicolumn{3}{@{\guiderule{oc-red-9}}l@{}}{
  \declkw{Inductive}~X : \mathcal{U}_p~\declkw{where}} \\
\multicolumn{3}{@{\guiderule{oc-red-9}}l@{}}{\quad \cdots} \\
  \quad \syn{c_{\kw{rel}}} & : & \mathbbm{2} \to X \\
  \quad \syn{c_{\kw{left}}} & : & \syn{c_{\kw{rel}}}~\mathsf{i0} = x \\
  \quad \syn{c_{\kw{right}}} & : & \syn{c_{\kw{rel}}}~\mathsf{i1} = x' .
\end{array}
\end{array}
\]
Thus no new primitive HIT mechanism is needed. The boxed reductions in
the inductive syntax, such as $\syn{\times_{\beta_1}}$, should be read in this way.
The same encoding of directed HITs as
ordinary HITs using the directed interval also appears in
\citet[Sections~4.4 and~5.3]{xue>2025}.

\subsubsection{Inductive Types in the Subuniverse of Preorders \AgdaFormalized}\label{sec:syntax:truncation}

The inductive types $\ssub$ and $\stm$ are valued in $\mathcal{U}_p$, the reflective subuniverse of preorders.
Within a reflective subuniverse, colimits (such as the inductive syntax here) are computed by applying the reflector $\mathsf{P} : \Univ \to \mathcal{U}_p$ (left adjoint to the inclusion $\mathcal{U}_p \hookrightarrow \Univ$, freely completing a type into a preorder) to the analogous colimit in $\Univ$~\cite{rijke>2019,rijke-shulman-spitters>2020}.
In particular, the schematic inductive definitions of $\ssub$ and $\stm$ computed in $\mathcal{U}_p$
are to be read as first defining raw HITs $\ssub_0~\Gamma~\Delta$ and
$\stm_0~\Gamma~A$ following
\cref{sec:syntax:directed-hit} and then applying the reflector:
\[
  \ssub~\Gamma~\Delta \isdef \mathsf{P}(\ssub_0~\Gamma~\Delta),
  \qquad
  \stm~\Gamma~A \isdef \mathsf{P}(\stm_0~\Gamma~A).
\]
The resulting types are preorders, whose universal properties within $\mathcal{U}_p$ follow from the universal property of $\mathsf{P}$ \citeappendix*{lem:localization-universal} and the universal properties of $\ssub_0$ and $\stm_0$ in $\Univ$.

\begin{theorem}[Initiality of the syntax \AgdaFormalized]
\label{thm:localized-syntax-initial}
The syntactic model, with substitution type $\ssub~\Gamma~\Delta$ and term
type $\stm~\Gamma~A$ as described, is initial among directed CwFs.
\end{theorem}

\subsubsection{Mapping Out of a Directed Quotient}\label{sec:syntax:mapping-out}

Elimination from a directed quotient $X$ needs to land in a preorder $Y : \mathcal{U}_p$, following the elimination
principle of reflective subuniverses~\cite[Lem.~2.17]{rijke-shulman-spitters>2020}. Defining a map $f : X \to Y$
requires an image for each point constructor of $X$ and, for each directed constructor $p : x \leq_X x'$, an inequality $f(p) :
f~x \leq_{Y} f~x'$. The latter is exactly the monotonicity condition.

\section{Logical Relations Model}\label{sec:gluing}

In this section, we construct a logical relations model of an object type theory with
directed reductions and use it to prove canonicity. In a gluing-style proof, such as that of \citet{kaposi-huber-sattler>2019}, the goal is to prove
a statement along the following lines: for every closed term
$M : \stm~\syn{\emp}~\syn{\kw{Bool}}$, either
$M = \syn{\kw{true}}$ or $M = \syn{\kw{false}}$, where equality is judgmental
equality in the object theory. Here our goal is instead to prove that $M$
reduces to either $\syn{\kw{true}}$ or $\syn{\kw{false}}$: in the language of simplicial type theory, that
$M \leq \syn{\kw{true}}$ or $M \leq \syn{\kw{false}}$.
The logical relations model is set in \glu{blue}.

The high-level strategy of proof-relevant logical relations is to
equip each type with two pieces of data, using a notation inspired by
\citet{sterling-harper>2021}: a syntactic component $A^\circ$, which
is the underlying syntactic type, and a semantic component $A^\bullet$, which
records evidence that terms of $A^\circ$ are well-behaved, or
computable. In particular, the semantic component $\glu{\kw{BOOL}}^\bullet$ of the Boolean type will imply the desired canonicity property.
We begin with non-dependent type formers.

\subsection{Semantics of the Judgmental Structure \AgdaFormalized}\label{sec:canonicity:judgmental}

We first recall the unary gluing model over global sections of the syntactic
CwF, ignoring directed reductions for now. Because canonicity concerns closed
terms, the semantic predicate $\Gamma^\bullet$ for a syntactic context $\Gamma^\circ$ is
indexed by closing substitutions $\gamma^\circ : \ssub~\syn{\emp}~\Gamma^\circ$. A substitution
$\sigma^\circ : \ssub~\Gamma^\circ~\Delta^\circ$ is computable when, composed
with computable global sections of $\Gamma^\circ$, it gives computable global
sections of $\Delta^\circ$.\footnote{We leave universe levels implicit throughout the paper.
For example, the record
$\glu{\kw{CTX}}$ must live at least one universe level higher than the universe in which $\Gamma^\bullet$ takes values.}
\[
\eqmakebox[glurecA][l]{$\begin{array}[t]{@{\guiderule{oc-blue-9}}l@{}}
  \declkw{record}~\glu{\kw{CTX}} : \Univ~\declkw{where} \\
  \quad \Gamma^\circ : \sctx \\
  \quad \Gamma^\bullet :
    \ssub~\syn{\emp}~\Gamma^\circ \to \Univ
\end{array}$}
\qquad
\eqmakebox[glurecB][l]{$\begin{array}[t]{@{\guiderule{oc-blue-9}}l@{}}
  \declkw{record}~\glu{\kw{SUB}}~
    (\Gamma~\Delta : \glu{\kw{CTX}}) : \Univ~\declkw{where} \\
  \quad \sigma^\circ :
    \ssub~\Gamma^\circ~\Delta^\circ \\
  \quad \sigma^\bullet :
    (\gamma^\circ : \ssub~\syn{\emp}~\Gamma^\circ) \to
    \Gamma^\bullet~\gamma^\circ \to
    \Delta^\bullet~(\sigma^\circ \scomp \gamma^\circ)
\end{array}$}
\]

A glued type has an underlying syntactic type $A^\circ : \sty~\Gamma^\circ$,
together with a computability predicate on closed terms of each closed instance
of $A^\circ$.
A glued term is then a syntactic term $M^\circ$ together with a computability
witness $M^\bullet$ that each closed instance of $M^\circ$ is computable at
$A^\bullet$.
\[
\eqmakebox[glurecA][l]{$\begin{array}[t]{@{\guiderule{oc-blue-9}}l@{}}
  \declkw{record}~\glu{\kw{TY}}~
    (\Gamma : \glu{\kw{CTX}}) : \Univ~\declkw{where} \\
  \quad A^\circ : \sty~\Gamma^\circ \\
  \quad A^\bullet :
    (\gamma^\circ : \ssub~\syn{\emp}~\Gamma^\circ) \to
    \Gamma^\bullet~\gamma^\circ \to \\
  \qquad
    \stm~\syn{\emp}~(\ssubst{A^\circ}{\gamma^\circ}) \to \Univ
\end{array}$}
\qquad
\eqmakebox[glurecB][l]{$\begin{array}[t]{@{\guiderule{oc-blue-9}}l@{}}
  \declkw{record}~\glu{\kw{TM}}~
    (\Gamma : \glu{\kw{CTX}})~
    (A : \glu{\kw{TY}}~\Gamma) : \Univ~\declkw{where} \\
  \quad M^\circ : \stm~\Gamma^\circ~A^\circ \\
  \quad M^\bullet :
    (\gamma^\circ : \ssub~\syn{\emp}~\Gamma^\circ) \to
    (\gamma^\bullet : \Gamma^\bullet~\gamma^\circ) \to \\
  \qquad
    A^\bullet~\gamma^\circ~\gamma^\bullet~
      (\ssubst{M^\circ}{\gamma^\circ})
\end{array}$}
\]

To see what extra structure is needed when we incorporate directed reductions, let us try to interpret the product
type. The proof for products will identify the missing ingredient: the
computability predicate of a type must be stable under directed expansion.
After that, we will return to the definition of $\glu{\kw{TY}}$ and strengthen
it accordingly.

\subsection{Semantics of Products \AgdaFormalized}\label{sec:canonicity:product}

As usual, a product is computable when both of its projections are computable.
\begin{logrel}
  \begin{array}[t]{@{}l@{}}
    \glu{\kw{PROD}} :
      \glu{\kw{TY}}~\Gamma \to
      \glu{\kw{TY}}~\Gamma \to
      \glu{\kw{TY}}~\Gamma
    \\[3pt]
    (\glu{\kw{PROD}}~A~B)^\circ
      \isdef A^\circ \mathbin{\syn{\times}} B^\circ
    \\
    (\glu{\kw{PROD}}~A~B)^\bullet~
      \gamma^\circ~\gamma^\bullet~P
      \isdef
      A^\bullet~\gamma^\circ~\gamma^\bullet~(\syn{\kw{fst}}~P)
      \times
      B^\bullet~\gamma^\circ~\gamma^\bullet~(\syn{\kw{snd}}~P)
  \end{array}
\end{logrel}
The projections are unchanged from the equational story: the syntactic
components are the syntactic projections, and the semantic components are the
metatheoretic projections.
\begin{logrel}
  \begin{array}[t]{@{}l@{}}
    \glu{\kw{FST}} :
      \glu{\kw{TM}}~\Gamma~(\glu{\kw{PROD}}~A~B) \to
      \glu{\kw{TM}}~\Gamma~A
    \\[3pt]
    (\glu{\kw{FST}}~P)^\circ
      \isdef \syn{\kw{fst}}~P^\circ
    \\
    (\glu{\kw{FST}}~P)^\bullet~
      \gamma^\circ~\gamma^\bullet
      \isdef \pi_1(P^\bullet~\gamma^\circ~\gamma^\bullet)
  \end{array}
  \qquad
  \begin{array}[t]{@{}l@{}}
    \glu{\kw{SND}} :
      \glu{\kw{TM}}~\Gamma~(\glu{\kw{PROD}}~A~B) \to
      \glu{\kw{TM}}~\Gamma~B
    \\[3pt]
    (\glu{\kw{SND}}~P)^\circ
      \isdef \syn{\kw{snd}}~P^\circ
    \\
    (\glu{\kw{SND}}~P)^\bullet~
      \gamma^\circ~\gamma^\bullet
      \isdef \pi_2(P^\bullet~\gamma^\circ~\gamma^\bullet).
  \end{array}
\end{logrel}
Now suppose
$M : \glu{\kw{TM}}~\Gamma~A$ and
$N : \glu{\kw{TM}}~\Gamma~B$. The syntactic component of
the pair is forced:
\begin{logrel}
  \begin{array}{@{}l@{}}
  \glu{\kw{PAIR}} :
    \glu{\kw{TM}}~\Gamma~A \to
    \glu{\kw{TM}}~\Gamma~B \to
    \glu{\kw{TM}}~\Gamma~(\glu{\kw{PROD}}~A~B)
  \\[3pt]
  (\glu{\kw{PAIR}}~M~N)^\circ
    \isdef \syn{\kw{pair}}~M^\circ~N^\circ .
  \end{array}
\end{logrel}
For the semantic component, after fixing
$\gamma^\circ : \ssub~\syn{\emp}~\Gamma^\circ$ and
$\gamma^\bullet : \Gamma^\bullet~\gamma^\circ$, the goal is to produce an
element of
$(\glu{\kw{PROD}}~A~B)^\bullet~\gamma^\circ~\gamma^\bullet~
(\ssubst{(\syn{\kw{pair}}~M^\circ~N^\circ)}{\gamma^\circ})$, which is a
metatheoretic product. The obvious definition would have the following shape:
\begin{logrel}
  \begin{array}[t]{@{}l@{\;}c@{\;}l@{}}
    \multicolumn{3}{@{}l@{}}{(\glu{\kw{PAIR}}~M~N)^\bullet~\gamma^\circ~\gamma^\bullet \isdef \Bigl(\hole{1}\;,\;\hole{2}\Bigr), \quad \text{where}} \\[3pt]
    \eqmakebox[pairName][l]{$\hole{1}$} & : &
      A^\bullet~\gamma^\circ~\gamma^\bullet~
        \bigl(\idx{\syn{\kw{fst}}~
          (\ssubst{(\syn{\kw{pair}}~M^\circ~N^\circ)}{\gamma^\circ})}\bigr) \\
    \eqmakebox[pairName][l]{$\hole{2}$} & : &
      B^\bullet~\gamma^\circ~\gamma^\bullet~
        \bigl(\idx{\syn{\kw{snd}}~
          (\ssubst{(\syn{\kw{pair}}~M^\circ~N^\circ)}{\gamma^\circ})}\bigr).
  \end{array}
\end{logrel}
But inductively the available witnesses are
\begin{logrel}
  \begin{array}[t]{@{}l@{\;}c@{\;}l@{}}
    \eqmakebox[pairName][l]{$M^\bullet~\gamma^\circ~\gamma^\bullet$} & : &
      A^\bullet~\gamma^\circ~\gamma^\bullet~
        \bigl(\idx{\ssubst{M^\circ}{\gamma^\circ}}\bigr) \\
    \eqmakebox[pairName][l]{$N^\bullet~\gamma^\circ~\gamma^\bullet$} & : &
      B^\bullet~\gamma^\circ~\gamma^\bullet~
        \bigl(\idx{\ssubst{N^\circ}{\gamma^\circ}}\bigr).
  \end{array}
\end{logrel}
In an equational proof-relevant logical relation, the two indices are identified
by $\beta$-equalities for products. In the directed syntax, however, these are
not equalities but merely reductions:
\begin{logrel}
  \begin{array}[t]{@{}l@{\;}c@{\;}l@{}}
    \eqmakebox[pairName][l]{$\ssubst{\syn{\times_{\beta_1}}}{\gamma^\circ}$} & : &
      \idx{\syn{\kw{fst}}~
        (\ssubst{(\syn{\kw{pair}}~M^\circ~N^\circ)}{\gamma^\circ})}
      \;\leq\; \idx{\ssubst{M^\circ}{\gamma^\circ}} \\
    \eqmakebox[pairName][l]{$\ssubst{\syn{\times_{\beta_2}}}{\gamma^\circ}$} & : &
      \idx{\syn{\kw{snd}}~
        (\ssubst{(\syn{\kw{pair}}~M^\circ~N^\circ)}{\gamma^\circ})}
      \;\leq\; \idx{\ssubst{N^\circ}{\gamma^\circ}}.
  \end{array}
\end{logrel}

In a reduction-based logical relation, one typically proves closure under (head)
expansion: if $T \rightarrow^*_\beta T'$ and $T'$ is computable, then so is $T$.
This suggests
equipping each semantic type with an operation transporting computability
witnesses backward along reductions $T \leq T'$ of closed terms:
\[
  \kw{expansion}_{\gamma^\circ,\gamma^\bullet} : T \leq T'
  \to
  A^\bullet~\gamma^\circ~\gamma^\bullet~
  T'
  \to
  A^\bullet~\gamma^\circ~\gamma^\bullet~
  T.
\]
This operation would solve the immediate typing problem above: it can turn
$M^\bullet~\gamma^\circ~\gamma^\bullet$ into a witness at the redex
$\syn{\kw{fst}}~(\ssubst{(\syn{\kw{pair}}~M^\circ~N^\circ)}{\gamma^\circ})$.
For proof-relevant predicates, however, the witness could be meaningful data,
so mere existence of backward transports is insufficient. At minimum, the transport
must be functorial in the sense that identity reductions induce identity maps
on witnesses, and expansion along a composite reduction is the composite of the
expansion maps.

\subsubsection{Contravariance as Proof-Relevant Expansion}
In simplicial type theory, contravariant families have been used as synthetic
analogues of right fibrations~\cite{riehl-shulman>2017} and to construct
directed univalent universes~\cite{gratzer-weinberger-buchholtz>2024,weaver-licata>2020,cavallo-riehl-sattler>2026}.
Here, contravariance provides exactly the right proof-relevant generalization
of the expansion lemma in logical relations and supplies the coherence needed
for expansion.
\begin{definition}[Contravariant family; {\protect\citealp[Def.~8.2]{riehl-shulman>2017}} \AgdaFormalized]
  A family $Y : X \to \Univ$ is
  \emph{contravariant} if for every $p : x \leq_X x'$ and $y' : Y(x')$,
  $\sum_{y : Y(x)} y \leq_{Y(p)} y'$ is contractible:
  \[
  \begin{array}{@{}l@{}}
    \kw{isContrav}~Y
    \isdef
    (x~x' : X)~
    (p : x \leq_X x')~
    (y' : Y(x')) \to
    \kw{isContr}~\Bigl(
      \sum_{y : Y(x)} y \leq_{Y(p)} y'
    \Bigr).
  \end{array}
  \]
\end{definition}
Every contravariant family $Y$ comes equipped with a backward transport operation.
\begin{definition}[Contravariant transport \AgdaFormalized]
  Suppose $Y : X \to \Univ$ is contravariant, witnessed by ${c_Y : \kw{isContrav}~Y}$. For $p : x \leq_X x'$, the
  \emph{contravariant transport} along $p$ is:
  \[
  \begin{array}{@{}l@{}}
    p^* : Y(x') \to Y(x) \\[3pt]
    p^*y'
      \isdef
      \pi_1\bigl(\kw{center}(c_Y~x~x'~p~y')\bigr).
  \end{array}
  \]
\end{definition}
Over a preordered base, contravariant transport is functorial and satisfies a
universal property.

\begin{lemma}[Functoriality of transport; {\protect\citealp[Prop.~8.16]{riehl-shulman>2017}} \AgdaFormalized]
  Suppose $X : \mathcal{U}_p$ is a preorder and $Y : X \to \Univ$ is contravariant.
  Contravariant transport preserves identities and composition: for
  $p : x \leq_X x'$, $q : x' \leq_X x''$, $y : Y(x)$, and $y'' : Y(x'')$,
  \[
    (\kw{id}_x)^*y = y
    \qquad\text{and}\qquad
    (p \cdot q)^*y'' = p^*(q^*y'').
  \]
\end{lemma}

\begin{lemma}[Universal property of transport; {\protect\citealp[Lem.~8.15]{riehl-shulman>2017}} \AgdaFormalized]\label{lem:contrav-universal}
  For a contravariant family $Y : X \to \Univ$ with $p : x \leq_X x'$,
  $y : Y(x)$, and $y' : Y(x')$, there is an equivalence
  \[
    (y \leq_{Y(p)} y') \simeq (y = p^*y').
  \]
\end{lemma}
The move is to impose this condition on every
computability predicate $A^\bullet$ carried by a glued type:
$
  \kw{isContrav}~(A^\bullet~\gamma^\circ~\gamma^\bullet)$.
The induced contravariant transport operation is the expansion operation needed
above. Thus a glued type stores not only its computability predicate,\footnote{The
predicates $\Gamma^\bullet$ and $A^\bullet$ are also required to be set-valued.
We omit this routine requirement from the displayed records to keep the
definitions focused on contravariance.} but also
the proof that this predicate is contravariant for each closed, computable substitution:
\[
\begin{array}[t]{@{\guiderule{oc-blue-9}}l@{}}
  \declkw{record}~\glu{\kw{TY}}~
    (\Gamma : \glu{\kw{CTX}}) : \Univ~\declkw{where} \\
  \quad A^\circ : \sty~\Gamma^\circ \\
  \quad A^\bullet :
    (\gamma^\circ : \ssub~\syn{\emp}~\Gamma^\circ) \to \\
  \qquad (\gamma^\bullet : \Gamma^\bullet~\gamma^\circ) \to
    \stm~\syn{\emp}~(\ssubst{A^\circ}{\gamma^\circ}) \to \Univ \\
  \quad
  \ineqbox{
    \begin{array}[t]{@{}l@{}}
    c_A : (\gamma^\circ : \ssub~\syn{\emp}~\Gamma^\circ) \to \\
    \quad (\gamma^\bullet : \Gamma^\bullet~\gamma^\circ) \to
      \kw{isContrav}~(A^\bullet~\gamma^\circ~\gamma^\bullet)
    \end{array}}
\end{array}
\qquad
\begin{array}[t]{@{\guiderule{oc-blue-9}}l@{}}
  \declkw{record}~\glu{\kw{TM}}~
    (\Gamma : \glu{\kw{CTX}})~
    (A : \glu{\kw{TY}}~\Gamma) : \ineqbox{\mathcal{U}_p}~\declkw{where} \\
  \quad M^\circ : \stm~\Gamma^\circ~A^\circ \\
  \quad M^\bullet :
    (\gamma^\circ : \ssub~\syn{\emp}~\Gamma^\circ) \to \\
  \qquad (\gamma^\bullet : \Gamma^\bullet~\gamma^\circ) \to
    A^\bullet~\gamma^\circ~\gamma^\bullet~
      (\ssubst{M^\circ}{\gamma^\circ})
\end{array}
\]

By \cref{sec:syntax:mapping-out}, $\glu{\kw{TM}}~\Gamma~A$ must be a
preorder to interpret the directed constructors of $\stm$, which is guaranteed by the following closure properties.

\begin{lemma}[Preorder total spaces of contravariant families \AgdaFormalized]
\label{lem:contrav-total-preorder}
  If $X : \mathcal{U}_p$ and $Y : X \to \Univ$ is contravariant and set-valued,
  then $\sum_{x:X} Y(x) : \mathcal{U}_p$.
\end{lemma}

\begin{corollary}[Glued substitutions and terms are preorders \AgdaFormalized]
\label{cor:glued-tm-preorder}
  For every $\Gamma,\Delta : \glu{\kw{CTX}}$ and $A : \glu{\kw{TY}}~\Gamma$,
  both $\glu{\kw{SUB}}~\Gamma~\Delta$ and $\glu{\kw{TM}}~\Gamma~A$
  lie in $\mathcal{U}_p$.
\end{corollary}

With the refined definitions of $\glu{\kw{TY}}$ and $\glu{\kw{TM}}$ in place,\footnote{Because a context is a product of types, contravariance of the predicates in $\glu{\kw{TY}}$ lifts to the context predicate $\Gamma^\bullet$ in $\glu{\kw{CTX}}$.}
inequalities between glued terms can be described explicitly. An inequality
$M \leq N$ between glued terms is an inequality in a $\Sigma$-type: it relates
both the underlying syntactic terms and the semantic witnesses.

By \cref{lem:sigma-hom}, an inequality $M \leq N$ at $\glu{\kw{TM}}~\Gamma~A$
consists of an underlying syntactic reduction
$\rho^\circ : M^\circ \leq N^\circ$, together with, for each closed
substitution $\gamma^\circ$ and computability witness $\gamma^\bullet$, a
dependent inequality between the two computability witnesses over the closed
reduction $\ssubst{\rho^\circ}{\gamma^\circ}$:
\[
  M^\bullet~\gamma^\circ~\gamma^\bullet
  \leq_{A^\bullet~\gamma^\circ~\gamma^\bullet~
    (\ssubst{\rho^\circ}{\gamma^\circ})}
  N^\bullet~\gamma^\circ~\gamma^\bullet .
\]
By \cref{lem:contrav-universal}, this dependent inequality is equivalently an
equality with the contravariant transport:
\[
  M^\bullet~\gamma^\circ~\gamma^\bullet
  =
  (\ssubst{\rho^\circ}{\gamma^\circ})^*
  (N^\bullet~\gamma^\circ~\gamma^\bullet).
\]
Thus the order on computability witnesses is forced by the contravariance
condition on the glued type: a syntactic reduction $M^\circ \leq N^\circ$ lifts to a glued inequality precisely when the witness for $M^\circ$ is obtained by transporting the witness for $N^\circ$
backward along that reduction.

\subsubsection{Glued Semantics for Product and Pair \AgdaFormalized}
Contravariance of the product predicate follows from that of its components by the following two closure properties.

\begin{lemma}[Contravariance under reindexing; {\protect\citealp[Rem.~8.3]{riehl-shulman>2017}} \AgdaFormalized]\label{lem:contrav-reindex}
  Let $f : X \to Y$ and $Z : Y \to \Univ$. If $Z$ is contravariant, then the
  reindexed family $Z \circ f$ is contravariant.
\end{lemma}

\begin{lemma}[Contravariance under products \AgdaFormalized]\label{lem:contrav-product}
  If $Y,Z : X \to \Univ$ are contravariant, then their pointwise product
  $x \mapsto Y(x) \times Z(x)$ is contravariant.
\end{lemma}

Reindex the contravariant families
$A^\bullet~\gamma^\circ~\gamma^\bullet$ and
$B^\bullet~\gamma^\circ~\gamma^\bullet$ along the projections
$\syn{\kw{fst}}$ and $\syn{\kw{snd}}$, respectively.
By \cref{lem:contrav-reindex}, both reindexed families are contravariant.
Their pointwise product is
$(\glu{\kw{PROD}}~A~B)^\bullet~\gamma^\circ~\gamma^\bullet$, so
\cref{lem:contrav-product} supplies $c_{\glu{\kw{PROD}}~A~B}$.

Finally, the semantic component of the pair can be defined:
\begin{logrel}
  \begin{array}{@{}l@{}}
    (\glu{\kw{PAIR}}~M~N)^\bullet~
      \gamma^\circ~\gamma^\bullet
      \isdef
      \Bigl(
        \ineqbox{(\ssubst{\syn{\times_{\beta_1}}}{\gamma^\circ})^*}~
          (M^\bullet~\gamma^\circ~\gamma^\bullet),\,
        \ineqbox{(\ssubst{\syn{\times_{\beta_2}}}{\gamma^\circ})^*}~
          (N^\bullet~\gamma^\circ~\gamma^\bullet)
      \Bigr).
  \end{array}
\end{logrel}
The transports are necessary because the product predicate is indexed by the
projections of the syntactic pair, while $M^\bullet$ and $N^\bullet$ live over
the two components themselves. The $\beta$-reductions bridge precisely this gap.
It remains to check that this interpretation respects the directed quotient
constructors for products. This is the semantic content of mapping out of a
directed quotient: the images of the point constructors
$\syn{\kw{pair}}$, $\syn{\kw{fst}}$, and $\syn{\kw{snd}}$ must
respect the inequalities
$\syn{\times_{\beta_1}}$ and $\syn{\times_{\beta_2}}$.

\subsubsection{Glued Terms Respect the Directed Quotient \AgdaFormalized}
First consider the inequality $\syn{\times_{\beta_1}}$ in the syntax. As explained in \cref{sec:syntax:mapping-out}, the mapping out (from syntactic terms) must respect this inequality. This is manifested as the following proof obligation:
\[
  \glu{\kw{FST}}~(\glu{\kw{PAIR}}~M~N)
  \leq_{\glu{\kw{TM}}~\Gamma~A}
  M .
\]
Because $\glu{\kw{TM}}$ is a $\Sigma$-type, by \cref{lem:sigma-hom}, this
inequality consists of two parts: a syntactic reduction and a dependent
inequality over it.
The syntactic part is exactly the product $\beta_1$-reduction:
\[
  (\glu{\kw{FST}}~(\glu{\kw{PAIR}}~M~N))^\circ
  =
  \syn{\kw{fst}}~(\syn{\kw{pair}}~M^\circ~N^\circ)
  \leq M^\circ .
\]
For the semantic part, fix
$\gamma^\circ : \ssub~\syn{\emp}~\Gamma^\circ$ and
$\gamma^\bullet : \Gamma^\bullet~\gamma^\circ$. The proof obligation is
\[
  (\glu{\kw{FST}}~(\glu{\kw{PAIR}}~M~N))^\bullet~
    \gamma^\circ~\gamma^\bullet
  \leq_{A^\bullet~\gamma^\circ~\gamma^\bullet~
    (\ssubst{\syn{\times_{\beta_1}}}{\gamma^\circ})}
  M^\bullet~\gamma^\circ~\gamma^\bullet .
\]
The left-hand side computes as follows:
\[
\begin{array}{@{}l@{\quad}l@{}}
  (\glu{\kw{FST}}~(\glu{\kw{PAIR}}~M~N))^\bullet~
    \gamma^\circ~\gamma^\bullet
  \\[3pt]
  =\; \pi_1\bigl((\glu{\kw{PAIR}}~M~N)^\bullet~
    \gamma^\circ~\gamma^\bullet\bigr)
  & \text{by definition of $\glu{\kw{FST}}^\bullet$}
  \\[3pt]
  =\; \pi_1\Bigl(
      (\ssubst{\syn{\times_{\beta_1}}}{\gamma^\circ})^*
        (M^\bullet~\gamma^\circ~\gamma^\bullet),\,
      (\ssubst{\syn{\times_{\beta_2}}}{\gamma^\circ})^*
        (N^\bullet~\gamma^\circ~\gamma^\bullet)
    \Bigr)
  & \text{by definition of $\glu{\kw{PAIR}}^\bullet$}
  \\[3pt]
  =\;
      (\ssubst{\syn{\times_{\beta_1}}}{\gamma^\circ})^*
        (M^\bullet~\gamma^\circ~\gamma^\bullet)
  & \text{by projection.}
\end{array}
\]
Therefore the required semantic inequality is
\[
  (\ssubst{\syn{\times_{\beta_1}}}{\gamma^\circ})^*
    (M^\bullet~\gamma^\circ~\gamma^\bullet)
  \leq_{A^\bullet~\gamma^\circ~\gamma^\bullet~
    (\ssubst{\syn{\times_{\beta_1}}}{\gamma^\circ})}
  M^\bullet~\gamma^\circ~\gamma^\bullet .
\]
By \cref{lem:contrav-universal}, this is equivalent to the reflexive equality
\[
  (\ssubst{\syn{\times_{\beta_1}}}{\gamma^\circ})^*
    (M^\bullet~\gamma^\circ~\gamma^\bullet)
  =
  (\ssubst{\syn{\times_{\beta_1}}}{\gamma^\circ})^*
    (M^\bullet~\gamma^\circ~\gamma^\bullet).
\]
This is precisely where the proof-relevant argument needs the universal property
of contravariant transport to supply the dependent inequality validating the
glued reduction.
The $\syn{\times_{\beta_2}}$ case is symmetric, using the second projection and
the contravariance of $B^\bullet$.

\subsubsection{What If We Have $\eta$-Reduction as Well? \AgdaFormalized}

The constructions above use only the product $\beta$-reductions; indeed, $\eta$-reductions are optional for canonicity.
If the syntax does include an
$\eta$-inequality, the present model validates the $\eta$-reduction:
${\syn{\times_{\eta}} : \syn{\kw{pair}}~(\syn{\kw{fst}}~P)~(\syn{\kw{snd}}~P) \leq P}$.
The reason is that projecting this $\eta$-reduction gives reductions with the
same endpoints as the corresponding $\beta$-reductions.
For
$P : \glu{\kw{TM}}~\Gamma~(\glu{\kw{PROD}}~A~B)$, the required glued reduction
is
\[
  \glu{\kw{PAIR}}~(\glu{\kw{FST}}~P)~(\glu{\kw{SND}}~P)
  \leq_{\glu{\kw{TM}}~\Gamma~(\glu{\kw{PROD}}~A~B)}
  P .
\]
For the semantic part, again fix $\gamma^\circ$ and $\gamma^\bullet$.
Unfolding the source gives
\[
\begin{array}{@{}l@{}}
  (\glu{\kw{PAIR}}~(\glu{\kw{FST}}~P)~(\glu{\kw{SND}}~P))^\bullet~
    \gamma^\circ~\gamma^\bullet
  =
  \Bigl(
    (\ssubst{\syn{\times_{\beta_1}}}{\gamma^\circ})^*
      \pi_1(P^\bullet~\gamma^\circ~\gamma^\bullet),\,
    (\ssubst{\syn{\times_{\beta_2}}}{\gamma^\circ})^*
      \pi_2(P^\bullet~\gamma^\circ~\gamma^\bullet)
  \Bigr).
\end{array}
\]
Unfolding $(\glu{\kw{PROD}}~A~B)^\bullet$ splits the required dependent inequality
over $\ssubst{\syn{\times_{\eta}}}{\gamma^\circ}$ into two components:
\[
\begin{array}{@{}l@{}}
  (\ssubst{\syn{\times_{\beta_1}}}{\gamma^\circ})^*
    \pi_1(P^\bullet~\gamma^\circ~\gamma^\bullet)
  \leq_{A^\bullet~\gamma^\circ~\gamma^\bullet~
    (\syn{\kw{fst}}~(\ssubst{\syn{\times_{\eta}}}{\gamma^\circ}))}
  \pi_1(P^\bullet~\gamma^\circ~\gamma^\bullet),
  \\[3pt]
  (\ssubst{\syn{\times_{\beta_2}}}{\gamma^\circ})^*
    \pi_2(P^\bullet~\gamma^\circ~\gamma^\bullet)
  \leq_{B^\bullet~\gamma^\circ~\gamma^\bullet~
    (\syn{\kw{snd}}~(\ssubst{\syn{\times_{\eta}}}{\gamma^\circ}))}
  \pi_2(P^\bullet~\gamma^\circ~\gamma^\bullet).
\end{array}
\]
By the universal property of contravariant transport, the first component is
equivalent to
\[
  (\ssubst{\syn{\times_{\beta_1}}}{\gamma^\circ})^*
    \pi_1(P^\bullet~\gamma^\circ~\gamma^\bullet)
  =
  (\syn{\kw{fst}}~(\ssubst{\syn{\times_{\eta}}}{\gamma^\circ}))^*
    \pi_1(P^\bullet~\gamma^\circ~\gamma^\bullet).
\]
Both $\ssubst{\syn{\times_{\beta_1}}}{\gamma^\circ}$ and
$\syn{\kw{fst}}~(\ssubst{\syn{\times_{\eta}}}{\gamma^\circ})$ are reductions from
$\ssubst{(\syn{\kw{fst}}~(\syn{\kw{pair}}~(\syn{\kw{fst}}~P^\circ)~
  (\syn{\kw{snd}}~P^\circ)))}{\gamma^\circ}$ to
$\ssubst{(\syn{\kw{fst}}~P^\circ)}{\gamma^\circ}$. The inequalities of the syntax are
propositions, so these reductions are equal, and hence their contravariant
transports agree. The second component follows by the same argument.

\subsection{Semantics of Functions \AgdaFormalized}\label{sec:canonicity:function}

The function type follows the same pattern as the product type. First recall
the syntax:
\[
\begin{array}{@{}c@{\hspace{3em}}c@{}}
\begin{array}[t]{@{}l@{\;}c@{\;}l@{}}
  -\mathbin{\syn{\Rightarrow}}- & : &
    \sty~\Gamma \to \sty~\Gamma \to \sty~\Gamma \\
  \syn{\kw{lam}} & : &
    \stm~(\Gamma \sext A)~(\ssubst{B}{\spp}) \to
    \stm~\Gamma~(A \mathbin{\syn{\Rightarrow}} B)
\end{array}
&
\begin{array}[t]{@{}l@{\;}c@{\;}l@{}}
  \syn{\kw{app}} & : &
    \stm~\Gamma~(A \mathbin{\syn{\Rightarrow}} B) \to
    \stm~\Gamma~A \to \stm~\Gamma~B \\
  \syn{\Rightarrow_{\beta}} & : &
    \ineqbox{
      \syn{\kw{app}}~(\syn{\kw{lam}}~N)~M
      \leq
      \ssubst{N}{\spair{\sidt}{M}}
    }.
\end{array}
\end{array}
\]
The logical relation for function types is standard: a computable
function is one that sends computable inputs to computable outputs.
\begin{logrel}
  \begin{array}{@{}l@{}}
    \glu{\kw{FUN}} :
      \glu{\kw{TY}}~\Gamma \to
      \glu{\kw{TY}}~\Gamma \to
      \glu{\kw{TY}}~\Gamma
    \\[3pt]
    (\glu{\kw{FUN}}~A~B)^\circ
      \isdef A^\circ \mathbin{\syn{\Rightarrow}} B^\circ
    \\
    (\glu{\kw{FUN}}~A~B)^\bullet~
      \gamma^\circ~\gamma^\bullet~F
      \isdef
      (M^\circ : \stm~\syn{\emp}~(\ssubst{A^\circ}{\gamma^\circ})) \to
      (M^\bullet : A^\bullet~\gamma^\circ~\gamma^\bullet~M^\circ) \to
      B^\bullet~\gamma^\circ~\gamma^\bullet~
        (\syn{\kw{app}}~F~M^\circ)
  \end{array}
\end{logrel}
To show contravariance for this predicate, we use the following closure
property.

\begin{lemma}[Contravariance under $\Pi$-types; {\protect\citealp[Thm.~8.30]{riehl-shulman>2017}} \AgdaFormalized]\label{lem:contrav-pi}
  Let $Z : X \to Y \to \Univ$. If $x \mapsto Z(x,y)$ is contravariant for
  every $y : Y$, then so is $x \mapsto ((y : Y) \to Z(x,y))$.
\end{lemma}

For each closed argument $M^\circ$
and witness $M^\bullet$, reindex $B^\bullet~\gamma^\circ~\gamma^\bullet$
along $F \mapsto \syn{\kw{app}}~F~M^\circ$.
By \cref{lem:contrav-reindex}, the resulting family is contravariant in $F$.
The domains of the two quantifiers in
$(\glu{\kw{FUN}}~A~B)^\bullet~\gamma^\circ~\gamma^\bullet$
do not depend on $F$, so applying \cref{lem:contrav-pi} twice supplies
$c_{\glu{\kw{FUN}}~A~B}$.

\subsubsection{Substitution Structure \AgdaFormalized}
The substitution and context extension operations used below follow the standard
gluing construction, and the full list is collected in \citeappendix{sec:reference-definitions}.
The only change from the equational presentation occurs in operations
involving contexts and types. For example, type substitution must also provide the
contravariance proof for $A\glu{[}\sigma\glu{]}$, inherited directly from the
one for $A$.

\subsubsection{Glued Semantics for Lambda and Application \AgdaFormalized}
The full constructor definitions are listed in \citeappendix{sec:reference-definitions}.
Only the semantic components are needed here:
\begin{logrel}
  \begin{array}{@{}l@{}}
    (\glu{\kw{LAM}}~N)^\bullet~
      \gamma^\circ~\gamma^\bullet~M^\circ~M^\bullet
      \isdef
      \ineqbox{(\ssubst{\syn{\Rightarrow_{\beta}}}
        {\spair{\gamma^\circ}{M^\circ}})^*}
      \bigl(
        N^\bullet~
          \spair{\gamma^\circ}{M^\circ}~
          (\gamma^\bullet,M^\bullet)
      \bigr)
    \\[8pt]
    (\glu{\kw{APP}}~F~M)^\bullet~
      \gamma^\circ~\gamma^\bullet
      \isdef
      F^\bullet~\gamma^\circ~\gamma^\bullet~
        (\ssubst{M^\circ}{\gamma^\circ})~
        (M^\bullet~\gamma^\circ~\gamma^\bullet).
  \end{array}
\end{logrel}
The boxed transport is induced by the instantiated $\beta$-reduction
\[
\begin{array}{@{}l@{}}
  \ssubst{\syn{\Rightarrow_{\beta}}}{\spair{\gamma^\circ}{M^\circ}}
  :
  \syn{\kw{app}}~
    (\ssubst{(\syn{\kw{lam}}~N^\circ)}{\gamma^\circ})~
    M^\circ
  \leq
  \ssubst{N^\circ}{\spair{\gamma^\circ}{M^\circ}} .
\end{array}
\]
Thus the instantiated evidence $N^\bullet~\spair{\gamma^\circ}{M^\circ}~
    (\gamma^\bullet,M^\bullet) :
  B^\bullet~\gamma^\circ~\gamma^\bullet~
    (\ssubst{N^\circ}{\spair{\gamma^\circ}{M^\circ}})$ is transported from evidence over the $\beta$-contractum to
evidence over the application of the lambda.

\subsubsection{Glued Functions Respect the Directed Quotient \AgdaFormalized}
Consider $\syn{\Rightarrow_{\beta}}$. For
$N : \glu{\kw{TM}}~(\Gamma \gext A)~(B\glu{[}\glu{\kw{p}}\glu{]})$ and
$M : \glu{\kw{TM}}~\Gamma~A$, the required glued inequality is
\[
  \glu{\kw{APP}}~(\glu{\kw{LAM}}~N)~M
  \leq_{\glu{\kw{TM}}~\Gamma~B}
  N\glu{[}\glu{(}\glu{\kw{ID}}\glu{,}M\glu{)}\glu{]} .
\]
For the semantic part of this inequality, fix $\gamma^\circ$ and $\gamma^\bullet$. The left-hand
side computes to
\[
\begin{array}{@{}l@{\quad}l@{}}
  (\glu{\kw{APP}}~(\glu{\kw{LAM}}~N)~M)^\bullet~
    \gamma^\circ~\gamma^\bullet
  \\[3pt]
  =\;
  (\glu{\kw{LAM}}~N)^\bullet~
    \gamma^\circ~\gamma^\bullet~
    (\ssubst{M^\circ}{\gamma^\circ})~
    (M^\bullet~\gamma^\circ~\gamma^\bullet)
  & \text{by definition of $\glu{\kw{APP}}^\bullet$}
  \\[3pt]
  =\;
  (\ssubst{\syn{\Rightarrow_{\beta}}}
    {\spair{\gamma^\circ}{\ssubst{M^\circ}{\gamma^\circ}}})^*
  \bigl(
    N^\bullet~
      \spair{\gamma^\circ}{\ssubst{M^\circ}{\gamma^\circ}}~
      (\gamma^\bullet,M^\bullet~\gamma^\circ~\gamma^\bullet)
  \bigr)
  & \text{by definition of $\glu{\kw{LAM}}^\bullet$.}
\end{array}
\]
The witness inside the transport is exactly the semantic component of the
substituted body:
\[
\begin{array}{@{}l@{}}
  (N\glu{[}\glu{(}\glu{\kw{ID}}\glu{,}M\glu{)}\glu{]})^\bullet~
    \gamma^\circ~\gamma^\bullet
  =
  N^\bullet~
    \spair{\gamma^\circ}{\ssubst{M^\circ}{\gamma^\circ}}~
    (\gamma^\bullet,M^\bullet~\gamma^\circ~\gamma^\bullet).
\end{array}
\]
By \cref{lem:contrav-universal}, the semantic inequality over
$\ssubst{\syn{\Rightarrow_{\beta}}}{\gamma^\circ}$ is equivalent to the
reflexive equality of this transported witness with itself, following the same
pattern as for products.

\subsection{Semantics of Booleans \AgdaFormalized}\label{sec:canonicity:bool}
The Boolean type is the point of the canonicity argument: its computability
predicate says that a closed term of Boolean type reduces to one of the two canonical Booleans. First recall the Boolean syntax with its non-dependent eliminator.
\[
\begin{array}{@{}c@{\hspace{3em}}c@{}}
\begin{array}[t]{@{}l@{\;}c@{\;}l@{}}
  \syn{\kw{Bool}} &:& \sty~\Gamma \\
  \syn{\kw{true}} &:& \stm~\Gamma~\syn{\kw{Bool}} \\
  \syn{\kw{false}} &:& \stm~\Gamma~\syn{\kw{Bool}}
\end{array}
&
\begin{array}[t]{@{}l@{\;}c@{\;}l@{}}
  \syn{\kw{if}}
  &:&
  \begin{array}[t]{@{}l@{}}
    (C : \sty~\Gamma) \to
    \stm~\Gamma~C \to
    \stm~\Gamma~C \to \\
    \stm~\Gamma~\syn{\kw{Bool}} \to
    \stm~\Gamma~C
  \end{array}
  \\
  \syn{\kw{Bool}_{\beta_1}}
  &:&
  \ineqbox{\syn{\kw{if}}~C~U~V~\syn{\kw{true}} \leq U}
  \\
  \syn{\kw{Bool}_{\beta_0}}
  &:&
  \ineqbox{\syn{\kw{if}}~C~U~V~\syn{\kw{false}} \leq V}.
\end{array}
\end{array}
\]
Write $\ulcorner 1 \urcorner \isdef \syn{\kw{true}}$ and $\ulcorner 0 \urcorner \isdef \syn{\kw{false}}$.
Then the glued Boolean type is:
\begin{logrel}
  \begin{array}{@{}l@{}}
    \glu{\kw{BOOL}} :
      \glu{\kw{TY}}~\Gamma
    \\[3pt]
    \glu{\kw{BOOL}}^\circ
      \isdef \syn{\kw{Bool}}
    \\
    \glu{\kw{BOOL}}^\bullet~
      \gamma^\circ~\gamma^\bullet~M
      \isdef
      \displaystyle
      \ineqbox{
        \sum_{b : \{0,1\}}
          M \leq_{\stm~\syn{\emp}~\syn{\kw{Bool}}}
          \ulcorner b \urcorner} .
  \end{array}
\end{logrel}
In the present setting, the
contravariance proof for $\glu{\kw{BOOL}}^\bullet$ is a familiar closure property available off the shelf from simplicial type theory:
representable families are contravariant.

\begin{lemma}[Contravariance of representables; {\protect\citealp[Prop.~8.13]{riehl-shulman>2017}} \AgdaFormalized]\label{lem:representable-contravariant}
  For any $X : \mathcal{U}_p$ and $x' : X$, the representable family
  $
    \bigl(x \mapsto x \leq_X x'\bigr) : X \to \Univ
  $
  is contravariant.
  Its contravariant transport sends
  $q : x'' \leq_X x'$ along $p : x \leq_X x''$ to the composite $p^*q \isdef p \cdot q : x \leq_X x'$.
\end{lemma}

For each $b : \{0,1\}$, the family
$M \mapsto M \leq \ulcorner b \urcorner$ is represented by
$\ulcorner b \urcorner$, so it is contravariant by \cref{lem:representable-contravariant};
because $\{0,1\}$ has no non-trivial inequalities (it is discrete, \cref{def:discrete}), the sum over $b$ is contravariant.
Immediately, the two canonical Booleans are computable:
\begin{logrel}
  \begin{array}{@{}l@{\qquad}l@{}}
    \begin{array}[t]{@{}l@{}}
      \glu{\kw{TRUE}} :
        \glu{\kw{TM}}~\Gamma~\glu{\kw{BOOL}}
      \\[3pt]
      \glu{\kw{TRUE}}^\circ
        \isdef \syn{\kw{true}}
      \\
      \glu{\kw{TRUE}}^\bullet~
        \gamma^\circ~\gamma^\bullet
        \isdef
        (1,\kw{id}_{\syn{\kw{true}}})
    \end{array}
    &
    \begin{array}[t]{@{}l@{}}
      \glu{\kw{FALSE}} :
        \glu{\kw{TM}}~\Gamma~\glu{\kw{BOOL}}
      \\[3pt]
      \glu{\kw{FALSE}}^\circ
        \isdef \syn{\kw{false}}
      \\
      \glu{\kw{FALSE}}^\bullet~
        \gamma^\circ~\gamma^\bullet
        \isdef
        (0,\kw{id}_{\syn{\kw{false}}}) .
    \end{array}
  \end{array}
\end{logrel}
The non-dependent eliminator is handled by case analysis on the Boolean computability
witness.
\begin{logrel}
  \begin{array}{@{}l@{}}
    \glu{\kw{IF}} :
      (C : \glu{\kw{TY}}~\Gamma) \to
      \glu{\kw{TM}}~\Gamma~C \to
      \glu{\kw{TM}}~\Gamma~C \to
      \glu{\kw{TM}}~\Gamma~\glu{\kw{BOOL}} \to
      \glu{\kw{TM}}~\Gamma~C
    \\[3pt]
    (\glu{\kw{IF}}~C~U~V~T)^\circ
      \isdef
      \syn{\kw{if}}~C^\circ~U^\circ~V^\circ~T^\circ .
  \end{array}
\end{logrel}
For the semantic component, fix $\gamma^\circ$ and $\gamma^\bullet$, and suppose
$T^\bullet~\gamma^\circ~\gamma^\bullet = (b,r)$. If $b=1$, set
\[
\begin{array}{@{}l@{}}
  \rho_1 :
  \syn{\kw{if}}~(\ssubst{C^\circ}{\gamma^\circ})~(\ssubst{U^\circ}{\gamma^\circ})~(\ssubst{V^\circ}{\gamma^\circ})~(\ssubst{T^\circ}{\gamma^\circ}) \leq
  \syn{\kw{if}}~(\ssubst{C^\circ}{\gamma^\circ})~(\ssubst{U^\circ}{\gamma^\circ})~(\ssubst{V^\circ}{\gamma^\circ})~\syn{\kw{true}}
  \leq
  \ssubst{U^\circ}{\gamma^\circ},
  \\[3pt]
  \rho_1
  \isdef
  \syn{\kw{if}}~(\ssubst{C^\circ}{\gamma^\circ})~(\ssubst{U^\circ}{\gamma^\circ})~(\ssubst{V^\circ}{\gamma^\circ})~r
  \cdot
  \ssubst{\syn{\kw{Bool}_{\beta_1}}}{\gamma^\circ}.
\end{array}
\]
If $b=0$, define $\rho_0$ in the same way, using
$\syn{\kw{Bool}_{\beta_0}}$ and ending at
$\ssubst{V^\circ}{\gamma^\circ}$. Then
\begin{logrel}
  \begin{array}{@{}l@{}}
    (\glu{\kw{IF}}~C~U~V~T)^\bullet~\gamma^\circ~\gamma^\bullet
    \isdef
    \begin{cases}
      \ineqbox{\rho_1^*}~(U^\bullet~\gamma^\circ~\gamma^\bullet),
        & \text{if } T^\bullet~\gamma^\circ~\gamma^\bullet = (1,r),\\[3pt]
      \ineqbox{\rho_0^*}~(V^\bullet~\gamma^\circ~\gamma^\bullet),
        & \text{if } T^\bullet~\gamma^\circ~\gamma^\bullet = (0,r).
    \end{cases}
  \end{array}
\end{logrel}
The two $\beta$-reductions $\glu{\kw{IF}}~C~U~V~\glu{\kw{TRUE}} \leq U$ and $\glu{\kw{IF}}~C~U~V~\glu{\kw{FALSE}} \leq V$ follow from the universal property of contravariant
transport, exactly as in the product and function cases.

\subsection{Canonicity via the Fundamental Theorem of Logical Relations \AgdaFormalized}\label{sec:canonicity:fundamental}
At this point, there are two CwF models: the initial syntactic model
$\smodel$ and the gluing model $\gmodel$. The gluing model $\gmodel$ has been constructed so that
each type $A^\circ$ of $\smodel$ is sent to $(A^\circ,A^\bullet)$ in
$\gmodel$, with the syntactic component $\circ$ in the gluing model exactly equal to
the corresponding component of the initial model. This is the fundamental theorem of
logical relations, shown in the diagram below.

For a syntactic closed term
$M : \stm~\syn{\emp}~\syn{\kw{Bool}}$, the glued term
$\interp{M}$ has syntactic projection $M$ itself. Its semantic projection,
instantiated at the closing substitution on the empty context, gives the witness displayed on the right.
The diagram on the left commutes by the initiality/induction principle of the
syntactic model.
\begingroup
\setlength{\abovedisplayskip}{2pt}
\setlength{\abovedisplayshortskip}{0pt}
\setlength{\belowdisplayskip}{0pt}
\setlength{\belowdisplayshortskip}{0pt}
\[
\begin{array}{@{}c@{\hspace{3em}}c@{}}
\begin{tikzcd}[ampersand replacement=\&, column sep=large, row sep=large]
  \smodel
    \arrow[r, "\interp{-}"]
    \arrow[dr, equals]
  \&
  \gmodel
    \arrow[d, "(-)^\circ"]
  \\
  \& \smodel
\end{tikzcd}
&
\begin{array}[c]{@{}l@{\;}c@{\;}l@{}}
  \interp{M}^\bullet~\seps_{\syn{\emp}}~()
  &:&
  \glu{\kw{BOOL}}^\bullet~\seps_{\syn{\emp}}~()~M
  \\[3pt]
  & = &
  \displaystyle\sum_{b : \{0,1\}}
      M \leq \ulcorner b \urcorner
\end{array}
\end{array}
\]
\endgroup
\begin{corollary}[Directed Boolean canonicity \AgdaFormalized]
For every closed term $M : \stm~\syn{\emp}~\syn{\kw{Bool}}$, there exists some $b : \{0,1\}$ with $M \leq \ulcorner b \urcorner$: every closed Boolean term reduces to either $\syn{\kw{true}}$ or $\syn{\kw{false}}$.
\end{corollary}

\section{Universes and Dependency}\label{sec:universe}

For the non-dependent type formers treated in \cref{sec:gluing}, adapting
proof-relevant logical relations from terms
quotiented by judgmental equality to terms quotiented by directed inequalities
requires only a modest strengthening of the induction hypothesis: each
computability predicate must be contravariant. Dependent type formers and universes add a further constraint: type conversion,
universe predicates, and dependency must remain compatible with contravariance.

\subsection{Obstacles to Naive Dependent Attempts}

Keeping the directed syntax from \cref{sec:syntax} and the contravariant
predicates from \cref{sec:gluing} leads to two related obstructions when adding
dependent type formers and a universe.

\subsubsection{How to Model Type Conversion?}

In ordinary type theory, type conversion is symmetric:
\[
  \inferrule{M : A \\ A \equiv B}{M : B}
\]
where $A \equiv B$ is judgmental equality of types. In the equational
logical relation, this rule is essentially invisible. Terms and types are
quotiented by judgmental equality, so judgmentally equal types are identified by
a path, and conversion becomes transport along that path in the metatheory:
\[
  \kw{transport}_{\stm~\Gamma} : A = B \to \stm~\Gamma~A \to \stm~\Gamma~B .
\]

This mechanism is not immediately available in the directed setting. The syntax records
type reductions as directed inequalities rather than judgmental equalities.
Quotienting first by judgmental equality would identify terms that the directed
quotient only relates by directed inequalities, thereby erasing the directed
structure. Thus the ordinary conversion rule has no premise of the required
shape: a reduction $A \leq_{\sty~\Gamma} B$ is not a judgmental equality
$A \equiv B$.

\subsubsection{Failure of the Naive Universe Predicate}
Proof-relevant logical relations allow a ``negative'' formulation of the
universe predicate: the computability of a code may itself be a
computability predicate for the decoded type. This formulation is unavailable
in proof-irrelevant settings where computability predicates land in
$\kw{Prop}$, because $\kw{Prop}$ cannot retain the required
data. In those settings, the universe predicate is usually formulated
``positively,'' as an inductive lookup table of type codes and their predicates,
restricting the universe to types specified in
advance~\cite{harper>1992,allen>thesis,angiuli>thesis,martin-lof>1972}.
The proof-relevant approach is more flexible. Consider a Tarski/Coquand-style
universe~\cite{luo>2012-universes,coquand>2018}, whose syntax is as follows:\footnote{Girard's paradox can be
avoided by considering a hierarchy of universes and a
hierarchy of judgments. This can be achieved by adding levels to $\sty$ and $\stm$
and lifting levels for the universe: $\syn{\kw{U}} : i \to
\sty_{i+1}~\Gamma$.}
\[
\begin{array}[t]{@{}l@{\;}c@{\;}l@{}}
  \syn{\kw{U}} &:& \sty~\Gamma \\
  \syn{\kw{U}_{\beta}} &:&
    \ineqbox{\syn{\kw{El}}~(\syn{\kw{Code}}~A) \leq A}
\end{array}
\qquad
\begin{array}[t]{@{}l@{\;}c@{\;}l@{}}
  \syn{\kw{El}} &:& \stm~\Gamma~\syn{\kw{U}} \to \sty~\Gamma \\
  \syn{\kw{Code}} &:& \sty~\Gamma \to \stm~\Gamma~\syn{\kw{U}} .
\end{array}
\]
The natural candidate for $\glu{\kw{UNIV}}$ records, at each closed code, a
computability predicate on the decoded type together with its contravariance
requirement, effectively internalizing $\glu{\kw{TY}}$:
\begin{logrel}
\begin{array}{@{}l@{}}
  \glu{\kw{UNIV}}^\circ \isdef \syn{\kw{U}} \\
  \glu{\kw{UNIV}}^\bullet~\gamma^\circ~\gamma^\bullet~(M : \stm~\syn{\emp}~\syn{\kw{U}})
    \isdef
    \sum_{P : \stm~\syn{\emp}~(\syn{\kw{El}}~M) \to \Univ}
      \kw{isContrav}~P .
\end{array}
\end{logrel}

This is the negative formulation: a code's computability content is itself a
glued type's worth of structure. The definition is attractive, but its
contravariant transport immediately runs into the same absence of type equality.
Given a universe reduction $r : M \leq_{\stm~\syn{\emp}~\syn{\kw{U}}} M'$ and a
predicate $P' : \stm~\syn{\emp}~(\syn{\kw{El}}~M') \to \Univ$, the decoded types
$\syn{\kw{El}}~M$ and $\syn{\kw{El}}~M'$ are related by a directed inequality,
not by a judgmental equality. Without equality between these decoded types,
there is no canonical transport between the corresponding fibers of $\stm$ on
which the predicates are defined.

Even if $P'$ could be transported backward to a predicate over
$\syn{\kw{El}}~M$, contravariance would require that transport to be unique, which is not guaranteed by
directed inequalities in a universe.
In summary,
the syntax calls for judgmental equality of types, the semantics for
equality of computability predicates. This section reconciles them
using \emph{discretization} and the \emph{flat modality}.

\subsection{Discrete Types and Flat Modality}\label{sec:universe:discrete}
The main idea is to work with judgmentally equivalent types while retaining the
directed reduction structure of terms. The critical observation is that
reductions are contained in judgmental equalities, except for symmetry. Freely
adding symmetry to directed structures in $\sty$ effectively
\emph{discretizes} $\sty~\Gamma$, making judgmentally equivalent types
available for conversion.

In simplicial type theory, a type is discrete when it has no non-trivial
directed inequalities.
\begin{definition}[Discrete type \AgdaFormalized]\label{def:discrete}
  A type $X$ is discrete if, for every $x,x':X$, the canonical map
  $x =_X x' \to x \leq_X x'$ is an equivalence:
  \[
    \kw{isDiscrete}~X \isdef
    (x~x':X) \to \kw{isEquiv}~(\_ : x =_{X} x' \to x \leq_X x').
  \]
\end{definition}

For example, the Boolean type $\{0,1\}$ is discrete. This is standard in
  simplicial type theory; see
  \citet[Axiom~6 and Corollary~3.20]{gratzer-weinberger-buchholtz>2024}.

Say a discrete type is a \emph{discrete set} when it is also a set.
Discrete sets form a reflective subuniverse $\mathcal{U}_s \hookrightarrow \mathcal{U}_p$; see \citeappendix{appendix:preorders:discrete}.
For the extension with universes and dependent types, we require syntactic
contexts and types to be discrete sets:
$\sctx:\mathcal{U}_s$ and $\sty~\Gamma:\mathcal{U}_s$.
Substitutions and terms remain in $\mathcal{U}_p$. As in
\cref{sec:syntax:truncation}, these conditions are imposed on the syntax
by reflecting the raw inductive syntax.

For a type reduction $r:A\leq_{\sty~\Gamma}B$, the discreteness of
$\sty~\Gamma$ gives a path $e_r:A=_{\sty~\Gamma}B$ by the inverse of the
canonical equivalence in \cref{def:discrete}. Transport along $e_r$ induces the equivalence
\[
  \kw{conv}_r
  \isdef
  \kw{transport}_{\stm~\Gamma}(e_r)
  : \stm~\Gamma~A \similarrightarrow \stm~\Gamma~B.
\]
We write $\kw{conv}$ when $r$ is clear from context.
Although $\sty~\Gamma$ is discrete, $\stm~\Gamma~\syn{\kw{U}}$ is not.
Reductions between universe codes remain directed, while $\syn{\kw{El}}$
sends them to equalities between decoded types.

\subsubsection{Contravariant Transport for the Universe}\label{sec:universe:universe-transport}
Given a universe reduction
$r : M \leq_{\stm~\syn{\emp}~\syn{\kw{U}}} M'$
and a contravariant predicate
$P' : \stm~\syn{\emp}~(\syn{\kw{El}}~M') \to \Univ$,
define the intended backward transport by
\[
  r^*P' \isdef P' \circ \kw{conv}_{\syn{\kw{El}}~r}.
\]
This predicate is contravariant by \cref{lem:contrav-reindex}.
Establishing contravariance of the universe predicate itself additionally
requires the unique lifting property. The following necessary condition
identifies the remaining obstruction.

\begin{lemma}[Fiberwise discreteness; {\protect\citealp[Prop.~8.18]{riehl-shulman>2017}} \AgdaFormalized]\label{lem:contrav-discrete-fibers}
  Let $Y : X \to \Univ$ be a contravariant family.
  Then $Y(x)$ is discrete for every $x : X$.
\end{lemma}

Applying this lemma to the proposed universe predicate requires discreteness
of each fiber
\[
  \sum_{P : \stm~\syn{\emp}~(\syn{\kw{El}}~M) \to \Univ}
    \kw{isContrav}~P.
\]
Contravariance of each $P$ ensures discreteness of its computability witnesses,
but does not ensure discreteness of this space of predicates. Directed
comparisons between predicates valued in the ambient universe $\Univ$ need
not identify them by equality.

At a high level, we want the universe predicate to classify ``the same''
computability predicates when syntactic universe codes are related by
reduction. This forces us to consider
a discrete space of predicates, where directed comparisons (between computability predicates) determine
equalities.

\subsubsection{Flat Modality}
Taking inspiration from crisp/modal type theory
\cite{shulman>2018-brouwer,licata-orton-pitts-spitters>2018,gratzer>thesis},
we store predicates under a discrete modality, so that inequalities between
computability predicates collapse to equalities. As
\citet{gratzer-weinberger-buchholtz>2024,gratzer-weinberger-buchholtz>2026}
show, simplicial type theory can be extended with modalities for synthetic
categorical reasoning. Among these is the flat modality $\flat$, which takes
the groupoid core of a type. From this point on, we work in this extension.
This modal extension is validated by the cubical-spaces model of
\citet[Section~4]{gratzer-weinberger-buchholtz>2024}.
For a type $X$, $\flat X$ is a discrete copy of $X$ obtained by removing
inequalities in $X$.
Because it discards structure, this modality is \emph{comonadic}.

Synthetically in type theory, this idempotent comonadic modality, analogous to a necessity modality in modal logic~\cite{pfenning-davies>2001}, requires a validity context $\Delta$ of flat variables and a normal context $\Gamma$. A term of type $\flat X$ can only be constructed using flat variables from the validity context~\cite[Figure~5]{shulman>2018-brouwer}:
\begin{mathpar}
  \inferrule*[right={$\flat$-form}]
    {\Delta~|~\emp \vdash X ~\kw{type}}
    {\Delta~|~\Gamma \vdash \flat X ~ \kw{type}}
  \and
  \inferrule*[right={$\flat$-intro}]
    {\Delta~|~\emp \vdash M : X}
    {\Delta~|~\Gamma \vdash M^\flat : \flat X}
  \and
  \inferrule*[right={$\flat$-elim}]
    {
      \Delta~|~\Gamma, x : \flat X \vdash C ~ \kw{type} \\
      \Delta~|~\Gamma \vdash M : \flat X \\
      \Delta, u : X~|~\Gamma \vdash N : C[u^\flat/x]
    }
    {
      \Delta~|~\Gamma \vdash
      \kw{let}^{\flat}~u = M~\kw{in}~N : C[M/x]
    }
\end{mathpar}
The associated comonadic counit is
\[
  \eps_X : \flat X \to X,
  \qquad
  \eps_X(x) \isdef \kw{let}^{\flat}~u = x~\kw{in}~u.
\]
It computes on introductions as $\eps_X(M^\flat) = M$. The counit gives the
following notion of flat modal type~\cite{rijke-shulman-spitters>2020}.
\begin{definition}\label{def:flat-modal}
  A (crisp) type is \emph{flat modal} when the comonadic counit is an equivalence:
  \[
    \kw{isFlatModal}~X \isdef \kw{isEquiv}~(\eps_X : \flat X \to X).
    \]
  In particular, $\flat X$ is flat modal by idempotence.
  For a flat modal type $X$, let $\kappa : X \to \flat X$ denote the inverse of the comonadic
  counit.
\end{definition}

\citet[Axiom~6]{gratzer-weinberger-buchholtz>2024} give the following characterization of flat modal types as an axiom: a (crisp) type is flat modal if and only if it is discrete.
Because $\flat X$ is flat modal, directed inequalities in $\flat X$ collapse to
equalities:
\[
  M \leq_{\flat X} N
  \quad\to\quad
  M =_{\flat X} N .
\]

In particular, if $X$ is discrete, hence flat modal, an element $x : X$
may be used as a flat variable in the validity context.

\subsubsection{Revised \texorpdfstring{$\glu{\kw{CTX}}$}{CTX} and \texorpdfstring{$\glu{\kw{TY}}$}{TY} with Flat Predicates}
We now construct $\glu{\kw{CTX}}$ and $\glu{\kw{TY}}$ in $\mathcal{U}_s$.
The idea is to store their computability predicates under the flat modality,
using types of the form
\[
  \flat(\ssub~\syn{\emp}~\Gamma^\circ \to \Univ)
  \qquad\text{and}\qquad
  \flat(\stm~\syn{\emp}~A^\circ \to \Univ).
\]
These types are discrete because every type of the form $\flat X$ is flat modal and hence discrete. Together with the
discrete syntactic fields and propositional contravariance fields, this
would make $\glu{\kw{CTX}}$ and $\glu{\kw{TY}}$ discrete. To form a model of our
set-truncated syntax (because $\sctx$ and $\sty$ are constructed in $\mathcal{U}_s$), however, the context and type records must also be
sets.

If the predicates are valued in the ambient univalent universe $\Univ$,
the type classifying such predicates need not be a set. This
is a common obstruction when doing type theory inside HoTT, as identified by \eg \citet[Section~6]{altenkirch-kaposi>2016} for the
standard model of set-truncated syntax. The logical relations model, being the \emph{gluing} of
the syntactic and standard models, also inherits this problem. Fortunately, prior work has identified several solutions to
obtain a hierarchy of strict universes that are themselves sets in HoTT. Examples include the inductive-recursive universe
of \citet{altenkirch-kaposi>2016}, the universe of sets of \citet[Section~10.5]{hott>2013}, and the universe of \citet{gratzer-gylterud-mortberg-stenholm>2024}.
We therefore ask that our computability predicates land in one such universe
$\mathcal{V}$ that is closed under enough types to construct computability predicates (\eg dependent products and sums,
equality and inequality types, Booleans, universes, and syntactic constructors such as $\stm$ and $\sty$); like $\Univ$, it need not be discrete.

By construction,
$\sty~\Gamma$ is discrete. The computability predicate for a type can therefore
be stored in $\glu{\kw{TY}}$ as a flat predicate. For a closed syntactic
type $A^\circ : \sty~\syn{\emp}$, define the type of flat predicates on
closed terms of $A^\circ$ by:
\[
  \Pred(A^\circ) \isdef \kw{let}^{\flat}~B = \kappa(A^\circ)~\kw{in}~
  \flat(\stm~\syn{\emp}~B \to \mathcal{V}).
\]

The glued context and type records are refined to store flat predicates alongside
the ordinary data. Substitutions and terms remain in $\mathcal{U}_p$ as in \cref{sec:gluing}.
\[
\begin{array}[t]{@{\guiderule{oc-blue-9}}l@{}}
  \declkw{record}~\glu{\kw{CTX}} : \ineqbox{\mathcal{U}_s}~\declkw{where} \\
  \quad \Gamma^\circ : \sctx \\
  \quad \ineqbox{\Gamma^\bullet_\flat :
    \flat(\ssub~\syn{\emp}~\Gamma^\circ \to \mathcal{V})} \\
  \quad \ineqbox{\Gamma^\bullet \isdef \eps(\Gamma^\bullet_\flat)} \\
  \quad c_\Gamma : \kw{isContrav}~\Gamma^\bullet
\end{array}
\qquad
\begin{array}[t]{@{\guiderule{oc-blue-9}}l@{}}
  \declkw{record}~\glu{\kw{TY}}~
    (\Gamma : \glu{\kw{CTX}}) : \ineqbox{\mathcal{U}_s}~\declkw{where} \\
  \quad A^\circ : \sty~\Gamma^\circ \\
  \quad \ineqbox{A^\bullet_\flat :
    (\gamma^\circ : \ssub~\syn{\emp}~\Gamma^\circ)
    \to (\gamma^\bullet : \Gamma^\bullet~\gamma^\circ)
    \to \Pred(\ssubst{A^\circ}{\gamma^\circ})} \\
  \quad \ineqbox{A^\bullet~\gamma^\circ~\gamma^\bullet
    \isdef \eps(A^\bullet_\flat~\gamma^\circ~\gamma^\bullet)} \\
  \quad c_A : (\gamma^\circ : \ssub~\syn{\emp}~\Gamma^\circ) \to
  (\gamma^\bullet : \Gamma^\bullet~\gamma^\circ) \to
    \kw{isContrav}~(A^\bullet~\gamma^\circ~\gamma^\bullet)
\end{array}
\]
The components $\Gamma^\bullet$ and $A^\bullet$ are now the counit-defined
views of the underlying flat predicates, with decoding from
$\mathcal{V}$ left implicit. The strict universe $\mathcal{V}$ supplies
setness, and the flat modality supplies discreteness. Therefore the revised records lie in
$\mathcal{U}_s$.

\subsubsection{Revisiting Products with Flat Predicates}

The construction from \cref{sec:gluing} can now be replayed with flat predicates. Products show the basic pattern. The syntactic component is unchanged,
while the semantic component is stored as a flat
predicate by
the functorial action of the flat modality.
\[
\begin{array}{@{}l@{}}
  (\glu{\kw{PROD}}~A~B)^\circ
    \isdef A^\circ \mathbin{\syn{\times}} B^\circ
  \\
  (\glu{\kw{PROD}}~A~B)^\bullet_\flat~\gamma^\circ~\gamma^\bullet
    \isdef
    \kw{let}^{\flat}~A^\bullet =
      A^\bullet_\flat~\gamma^\circ~\gamma^\bullet~\kw{in}~
    \kw{let}^{\flat}~B^\bullet =
      B^\bullet_\flat~\gamma^\circ~\gamma^\bullet~\kw{in}~
    \bigl(
      \lambda P.\;
        A^\bullet~(\syn{\kw{fst}}~P)
        \times
        B^\bullet~(\syn{\kw{snd}}~P)
    \bigr)^\flat .
\end{array}
\]
The predicate used in \cref{sec:canonicity:product} is recovered by
opening this flat predicate with the counit:
\[
  (\glu{\kw{PROD}}~A~B)^\bullet~\gamma^\circ~\gamma^\bullet
  \isdef
  \eps\bigl((\glu{\kw{PROD}}~A~B)^\bullet_\flat~\gamma^\circ~\gamma^\bullet\bigr).
\]
Because the counit computes through the flat eliminations, the recovered predicate is exactly the product predicate from
\cref{sec:canonicity:product}:
\[
  (\glu{\kw{PROD}}~A~B)^\bullet~\gamma^\circ~\gamma^\bullet~P
  =
  A^\bullet~\gamma^\circ~\gamma^\bullet~(\syn{\kw{fst}}~P)
  \times
  B^\bullet~\gamma^\circ~\gamma^\bullet~(\syn{\kw{snd}}~P).
\]
Thus storing computability predicates flatly does not change the predicate seen
by terms. Glued substitutions and terms retain the definitions of \cref{sec:gluing},
as $\glu{\kw{SUB}}$ and $\glu{\kw{TM}}$ refer only to the counit-applied
predicates
$\Gamma^\bullet \isdef \eps(\Gamma^\bullet_\flat)$ and
$A^\bullet~\gamma^\circ~\gamma^\bullet
    \isdef \eps(A^\bullet_\flat~\gamma^\circ~\gamma^\bullet)$.
Contravariant transports are handled through the same counit-defined views.

\subsection{The Universe Glued Type}

With the infrastructure of the flat modality, we can finally define $\glu{\kw{UNIV}}$.
For each closed syntactic type $A^\circ : \sty~\syn{\emp}$, a candidate
consists of a flat predicate together with a proof that its recovered
predicate is contravariant:
\[
  \Cand(A^\circ)
    \isdef
    \sum_{A^\bullet_\flat : \Pred(A^\circ)}
      \kw{isContrav}~(\eps(A^\bullet_\flat)).
\]
The predicate assigns to each closed code
$M : \stm~\syn{\emp}~\syn{\kw{U}}$ the candidates on its decoded type
$\syn{\kw{El}}~M$:
\begin{logrel}
\begin{array}{@{}l@{}}
  \glu{\kw{UNIV}}^\circ
    \isdef \syn{\kw{U}},
  \\
  \glu{\kw{UNIV}}^\bullet_\flat~\gamma^\circ~\gamma^\bullet
    \isdef
    \bigl(\lambda M.\; \Cand(\syn{\kw{El}}~M)\bigr)^\flat.
\end{array}
\end{logrel}

The outer flat stores this family in the form required by
$\glu{\kw{TY}}$, while the inner flat makes comparisons of predicates discrete. Contravariance follows from a general fact about discrete
families.

\begin{lemma}[Contravariance over a discrete base \AgdaFormalized]\label{lem:contrav-discrete-base}
  If $X$ is discrete and $Y : X \to \Univ$ has discrete fibers, then $Y$ is
  contravariant.
\end{lemma}

For each $A^\circ$, $\Pred(A^\circ)$ is discrete by
flat discreteness~\citep[Axiom~6]{gratzer-weinberger-buchholtz>2024}, and $\kw{isContrav}~(\eps(A^\bullet_\flat))$ is a
proposition. Therefore $\Cand(A^\circ)$ is discrete. Because $\sty~\syn{\emp}$ is
discrete by construction, \cref{lem:contrav-discrete-base} shows that $\Cand$ is a
contravariant family. The recovered universe
predicate is exactly $\Cand \circ \syn{\kw{El}}$, so
\cref{lem:contrav-reindex} establishes $c_{\glu{\kw{UNIV}}}$.

\subsection{El and Code}

The remaining Tarski operations are mostly bookkeeping. For
$M : \glu{\kw{TM}}~\Gamma~\glu{\kw{UNIV}}$,
$M^\bullet~\gamma^\circ~\gamma^\bullet$ is a candidate in
$\Cand\bigl(\syn{\kw{El}}~(\ssubst{M^\circ}{\gamma^\circ})\bigr)$, whose two
components $\glu{\kw{EL}}$ projects out. In the
other direction, if $A : \glu{\kw{TY}}~\Gamma$, then $\glu{\kw{CODE}}~A$
produces a candidate at
$\Cand\bigl(\syn{\kw{El}}~(\ssubst{(\syn{\kw{Code}}~A^\circ)}{\gamma^\circ})\bigr)$ by transporting the predicate and contravariance proof of $A$ along the reduction
$\ssubst{\syn{\kw{U}_{\beta}}}{\gamma^\circ}
  : \syn{\kw{El}}~(\ssubst{(\syn{\kw{Code}}~A^\circ)}{\gamma^\circ})
    \leq \ssubst{A^\circ}{\gamma^\circ}$.
\begin{logrel}
\begin{array}{@{}l@{\qquad}l@{}}
  \begin{array}{@{}l@{\;}c@{\;}l@{}}
    (\glu{\kw{EL}}~M)^\circ
      &\isdef& \syn{\kw{El}}~M^\circ
    \\
    (\glu{\kw{EL}}~M)^\bullet_\flat~\gamma^\circ~\gamma^\bullet
      &\isdef& \pi_1(M^\bullet~\gamma^\circ~\gamma^\bullet)
    \\
    c_{\glu{\kw{EL}}~M}~\gamma^\circ~\gamma^\bullet
      &\isdef& \pi_2(M^\bullet~\gamma^\circ~\gamma^\bullet).
  \end{array}
  &
  \begin{array}{@{}l@{\;}c@{\;}l@{}}
    (\glu{\kw{CODE}}~A)^\circ
      &\isdef& \syn{\kw{Code}}~A^\circ
    \\
    (\glu{\kw{CODE}}~A)^\bullet~
      \gamma^\circ~\gamma^\bullet
      &\isdef&
      \ineqbox{(\ssubst{\syn{\kw{U}_{\beta}}}{\gamma^\circ})^*}
        \bigl(A^\bullet_\flat~\gamma^\circ~\gamma^\bullet,\,
              c_A~\gamma^\circ~\gamma^\bullet\bigr).
  \end{array}
\end{array}
\end{logrel}

The constructions for dependent type formers follow the same pattern as in
\cref{sec:gluing}, using discreteness of types to handle type
conversion. Their definitions are collected in \citeappendix{sec:reference-definitions}.
Together with the universe construction,
they extend the gluing model and yield Boolean canonicity by the same argument of \cref{sec:canonicity:fundamental}.
\section{Binary Logical Relations and Parametricity}\label{sec:binary}

The unary model proves canonicity by attaching to each closed term a
computability witness. For parametricity, the binary model has two
syntactic terms, as in the doubled syntax of
\citet[Computation~5.29]{sterling-harper>2021}. Its semantic component is computability evidence that relates
closed instances of the left and right syntactic terms. The proof-relevant
universe can therefore contain
contravariant, heterogeneous relations, supporting representation-independence proofs.

\begin{lemma}[Binary contravariance; {\protect\citealp[Prop.~8.21]{riehl-shulman>2017}} \AgdaFormalized]\label{lem:binary-contravariance}
  A binary family $R : X \to Y \to \Univ$
  is contravariant if and only if
  $R~x~(-) : Y \to \Univ$ and $R~(-)~y : X \to \Univ$ are both contravariant for each $x : X$ and $y : Y$, respectively.
\end{lemma}
Ordinary binary logical relations phrase closure under expansion one component
at a time:
\[
\begin{array}{@{}l@{}}
  \text{if } M_L \rightarrow_\beta M'_L,
  \text{then } R~M'_L~M_R
  \to
  R~M_L~M_R,
  \qquad
  \text{if } M_R \rightarrow_\beta M'_R,
  \text{then } R~M_L~M'_R
  \to
  R~M_L~M_R.
\end{array}
\]
\Cref{lem:binary-contravariance} identifies the directed version of these two
principles with contravariance of the uncurried relation. Thus reductions
$r_L : M_L \leq M'_L$ and $r_R : M_R \leq M'_R$ determine a transport
\[
  (r_L,r_R)^* : R~M'_L~M'_R \to R~M_L~M_R .
\]
This transport also records functoriality and its
universal property. As in \cref{sec:universe}, the relation carried by a type must be stored as a
flat predicate. For closed types
$A_L^\circ,A_R^\circ : \sty~\syn{\emp}$, write
\[
\begin{array}{@{}l@{}}
  \Pred_2(A_L^\circ,A_R^\circ)
  \isdef
  \kw{let}^{\flat}~B_L = \kappa(A_L^\circ)~\kw{in}~
  \kw{let}^{\flat}~B_R = \kappa(A_R^\circ)~\kw{in}~
  \flat(\stm~\syn{\emp}~B_L \to \stm~\syn{\emp}~B_R \to \mathcal{V}),
  \\[8pt]
  \Cand_2(A_L^\circ,A_R^\circ)
  \isdef
  \sum_{A^\bullet_\flat : \Pred_2(A_L^\circ,A_R^\circ)}
    \kw{isContrav}~(\eps(A^\bullet_\flat)).
\end{array}
\]
A glued context relates closing substitutions into its left and right contexts.
A glued type assigns a predicate that relates the pair of terms from the left and right syntactic types.
\[
\begin{array}{@{}l@{\qquad}l@{}}
  \begin{array}[t]{@{\guiderule{oc-blue-9}}l@{}}
    \declkw{record}~\glu{\kw{CTX}_2} : \mathcal{U}_s~\declkw{where} \\
    \quad \Gamma_L^\circ,\,\Gamma_R^\circ : \sctx \\
    \quad \Gamma^\bullet_\flat :
      \flat(\ssub~\syn{\emp}~\Gamma_L^\circ
        \to \ssub~\syn{\emp}~\Gamma_R^\circ \to \mathcal{V}) \\
    \quad \Gamma^\bullet \isdef \eps(\Gamma^\bullet_\flat) \\
    \quad c_\Gamma : \kw{isContrav}~\Gamma^\bullet
  \end{array}
  &
  \begin{array}[t]{@{\guiderule{oc-blue-9}}l@{}}
    \declkw{record}~\glu{\kw{TY}_2}~
      (\Gamma : \glu{\kw{CTX}_2}) : \mathcal{U}_s~\declkw{where} \\
    \quad A_L^\circ : \sty~\Gamma_L^\circ \qquad A_R^\circ : \sty~\Gamma_R^\circ \\
    \quad A^\bullet_\flat :
      (\gamma_L^\circ : \ssub~\syn{\emp}~\Gamma_L^\circ)
      \to (\gamma_R^\circ : \ssub~\syn{\emp}~\Gamma_R^\circ) \\
    \qquad \to \Gamma^\bullet~\gamma_L^\circ~\gamma_R^\circ \\
    \qquad
      \to \ineqbox{\Pred_2(\ssubst{A_L^\circ}{\gamma_L^\circ},
                  \ssubst{A_R^\circ}{\gamma_R^\circ})} \\
    \quad A^\bullet~\gamma_L^\circ~\gamma_R^\circ~\gamma^\bullet
      \isdef \eps(A^\bullet_\flat~\gamma_L^\circ~\gamma_R^\circ~\gamma^\bullet) \\
    \quad c_A : \forall\,\gamma_L^\circ~\gamma_R^\circ~\gamma^\bullet.\;
      \kw{isContrav}~
        (A^\bullet~\gamma_L^\circ~\gamma_R^\circ~\gamma^\bullet).
  \end{array}
\end{array}
\]
A glued term has a term on each side and a witness relating their closed instances.
\[
\begin{array}{@{\guiderule{oc-blue-9}}l@{}}
  \declkw{record}~\glu{\kw{TM}_2}~
    (\Gamma : \glu{\kw{CTX}_2})~
    (A : \glu{\kw{TY}_2}~\Gamma) : \mathcal{U}_p~\declkw{where} \\
  \quad M_L^\circ : \stm~\Gamma_L^\circ~A_L^\circ \qquad
        M_R^\circ : \stm~\Gamma_R^\circ~A_R^\circ \\
  \quad M^\bullet :
    (\gamma_L^\circ : \ssub~\syn{\emp}~\Gamma_L^\circ)
    \to (\gamma_R^\circ : \ssub~\syn{\emp}~\Gamma_R^\circ)
    \to (\gamma^\bullet :
      \Gamma^\bullet~\gamma_L^\circ~\gamma_R^\circ) \\
  \qquad \to \ineqbox{A^\bullet~\gamma_L^\circ~\gamma_R^\circ~\gamma^\bullet~
      (\ssubst{M_L^\circ}{\gamma_L^\circ})~(\ssubst{M_R^\circ}{\gamma_R^\circ})}.
\end{array}
\]

As in \cref{sec:canonicity:fundamental}, the gluing construction forms a CwF $\gmodel_2$, and the two syntactic projections give a CwF morphism
$(-)_L^\circ\times(-)_R^\circ:\gmodel_2\to\smodel\times\smodel$. For the interpretation
$\interp{-}:\smodel\to\gmodel_2$, the composite $((-)_L^\circ\times(-)_R^\circ)\circ\interp{-}$ is the
diagonal because each projection of $\interp{-}$ is the identity by initiality.
Thus the fundamental theorem interprets a syntactic term $M^\circ$ with projections
$(M^\circ,M^\circ)$ and relates $\ssubst{M^\circ}{\gamma_L^\circ}$ to
$\ssubst{M^\circ}{\gamma_R^\circ}$ whenever the closing substitutions are related.

The binary universe is obtained by replacing unary candidates by binary
candidates:
\begin{logrel}
\begin{array}{@{}l@{}}
  \glu{\kw{UNIV}_2}^\bullet_\flat~
    \gamma_L^\circ~\gamma_R^\circ~\gamma^\bullet
  \isdef
    \bigl(\lambda M_L~M_R.\;
      \Cand_2(\syn{\kw{El}}~M_L,\syn{\kw{El}}~M_R)\bigr)^\flat .
\end{array}
\end{logrel}
As an example, consider the batched queue representation-independence argument
of \citet{sterling-harper>2021}. The abstract package is a queue interface with
a representation type, an empty queue, and enqueue and dequeue operations:
\[
\begin{array}{@{}l@{\;}c@{\;}l@{}}
  \kw{Queue} &\isdef&
    \stm~\syn{\emp}~\Bigl(
      \syn{\Sigma}~\syn{\kw{U}}~
        \bigl(
          \syn{\kw{El}}~\sqq
          \mathbin{\syn{\times}}
          (\syn{\kw{Nat}} \mathbin{\syn{\Rightarrow}}
            \syn{\kw{El}}~\sqq \mathbin{\syn{\Rightarrow}} \syn{\kw{El}}~\sqq)
          \mathbin{\syn{\times}}
          (\syn{\kw{El}}~\sqq \mathbin{\syn{\Rightarrow}}
            \syn{\kw{Nat}} \mathbin{\syn{\times}} \syn{\kw{El}}~\sqq)
        \bigr)
    \Bigr).
  \\[8pt]
  \kw{QUEUE} &\isdef&
    \glu{\kw{TM}_2}~\glu{\emp}~\Bigl(
      \glu{\Sigma}~\glu{\kw{UNIV}_2}~
        \bigl(
          \glu{\kw{EL}}~\glu{\kw{q}}
          \mathbin{\glu{\times}}
          (\glu{\kw{NAT}} \mathbin{\glu{\Rightarrow}}
            \glu{\kw{EL}}~\glu{\kw{q}} \mathbin{\glu{\Rightarrow}} \glu{\kw{EL}}~\glu{\kw{q}})
          \mathbin{\glu{\times}}
          (\glu{\kw{EL}}~\glu{\kw{q}} \mathbin{\glu{\Rightarrow}}
            \glu{\kw{NAT}} \mathbin{\glu{\times}} \glu{\kw{EL}}~\glu{\kw{q}})
        \bigr)
    \Bigr).
\end{array}
\]
For representation independence, we seek an element of $\kw{QUEUE}$ whose two
syntactic projections are the simple list queue and the batched queue,
respectively. Both are elements of $\kw{Queue}$. At the $\glu{\kw{UNIV}_2}$
component, choose $\alpha_L$ as the code of $\syn{\kw{List}~\kw{Nat}}$ for the
simple list queue, and $\alpha_R$ as the code of
$\syn{\kw{List}~\kw{Nat} \mathbin{\times} \kw{List}~\kw{Nat}}$ for the batched queue.
The abstraction relation has type
\[
  R :
  \stm~\syn{\emp}~(\syn{\kw{El}}~\alpha_L)
  \to
  \stm~\syn{\emp}~(\syn{\kw{El}}~\alpha_R)
  \to \mathcal{V}.
\]
\citet{sterling-harper>2021} express the extensional queue invariant using meta-level lists $\kw{List}~\mathbb{N}$:
\[
  R~M_L~M_R \isdef
  \sum_{F,B:\kw{List}~\mathbb{N}}~
    \bigl(M_L = \ulcorner \kw{append}~F~(\kw{reverse}~B)\urcorner\bigr)
    \times \bigl(\syn{\kw{fst}}~M_R = \ulcorner F\urcorner\bigr)
    \times \bigl(\syn{\kw{snd}}~M_R = \ulcorner B\urcorner\bigr).
\]
This relation is, however, not contravariant in the present setting. Instead,
we use the directed analogue below, which replaces each equality by a reduction to the
canonical form determined by a common pair of meta-level lists $F,B: \kw{List}~\mathbb{N}$.
For each such pair, the inequalities form a product of
representable families (the last two reindexed along $\syn{\kw{fst}}$ and
$\syn{\kw{snd}}$, respectively, by \cref{lem:contrav-reindex}), so contravariance is supplied by
\cref{lem:representable-contravariant}.
\[
  R~M_L~M_R \isdef
  \sum_{F,B:\kw{List}~\mathbb{N}}~
    \bigl(M_L \leq \ulcorner \kw{append}~F~(\kw{reverse}~B)\urcorner\bigr)
    \times \bigl(\syn{\kw{fst}}~M_R \leq \ulcorner F\urcorner\bigr)
    \times \bigl(\syn{\kw{snd}}~M_R \leq \ulcorner B\urcorner\bigr).
\]
\section{Conclusion, Discussion, Related and Future Work}\label{sec:conclusion}

The result proved in this paper is, in one sense, unsurprising: logical
relations should validate canonicity and parametricity when syntax is
presented by reductions. This paper shows how to internalize these arguments
synthetically in simplicial type theory. The choice of meta-language matters:
inequality types provide an account of reduction, while
contravariant families supply the proof-relevant form of the expansion lemma
needed by logical relations. Existing results about contravariance in simplicial type theory therefore serve as a
``standard library'' for establishing closure under expansion compositionally.
The construction also gives an application of simplicial type theory beyond
synthetic $\infty$-category theory. We now discuss related work and future directions.

\subsection{Internal and External Reduction}
A conventional inductively defined reduction relation, after taking its
reflexive--transitive closure, organizes terms into an external preorder.
A preorder is a thin category, making it suitable for synthetic reasoning in
simplicial type theory. Internally, our directed syntax is characterized by
its universal property as the initial directed CwF with the specified
constructors, equations, and reductions. This follows the familiar approach
of synthetic homotopy theory: just as the circle in HoTT is characterized
synthetically by the universal property expressed by its induction principle,
our syntax is characterized by its initiality with appropriate reductions (morphisms in a preorder).

How does this internally defined synthetic preorder relate to the external
preorder specified by reduction? The work of
\citet[Theorems~4.26 and~4.28]{grodin-niu-sterling-harper>2024}, developed
in a related simplicial setting, provides a starting point. That work
establishes soundness and completeness of synthetic inequalities on the embedding
of an ordinary preorder with respect to its order. Their results may therefore apply
to our setting if we can identify our directed syntax in the model of simplicial type theory
with the embedding of the conventional reduction preorder, perhaps by showing
that they satisfy the same universal property. Such a comparison would allow
internal canonicity to be read as an ordinary reduction result. We leave this
comparison to future work.

\subsection{Mechanization}
Mechanizing gluing and proof-relevant logical relations with intrinsic syntax,
as in the present work, is difficult mainly because of \emph{transport}. Many
equations are not definitional, such as naturality for pairs
${\ssubst{(\syn{\kw{pair}}~M~N)}{\sigma}
  = \syn{\kw{pair}}~(\ssubst{M}{\sigma})~(\ssubst{N}{\sigma})}$; here they arise from path constructors in a higher
inductive type. A mechanization must therefore transport along paths and reason about equalities (and, in the
present setting, inequalities) between transported terms. Existing approaches
either use extensional proof assistants with equality reflection such as NuPRL-style systems~\cite{constable-etal>1986}, as in
\citet{li-yao-harper>2026}, or encode the syntax in specific ways so that these equations become
definitional~\cite{kaposi-pujet>2025,xie-bense>2026} using variants of
observational type theory~\cite{altenkirch-mcbride-swierstra>2007,pujet-tabareau>2022}. The present work is committed to homotopy type theory as
a meta-language, so neither move is available. Our experience matches that of
\citet{chen-nordvall-forsberg-tsai>2026}: formalizing intrinsic syntax in
Cubical Agda remains challenging.

We provide a Cubical Agda mechanization of the fragment with non-dependent type
formers in \cref{sec:syntax,sec:gluing}. It formalizes directed inequalities,
contravariance, and intrinsic syntax with products, functions, and Booleans. The resulting gluing model yields Boolean canonicity. Cubical Agda is used because it provides higher inductive types and a
higher-dimensional meta-language. As in other work on simplicial type theory
in Cubical Agda~\cite{de-jong-kraus-ljungstrom>2026,chhabra-angiuli-gratzer>2026},
our mechanization postulates a directed interval and its axioms. Even with
non-dependent type formers, the transport burden is already somewhat unpleasant, though
manageable with care. The extension to dependent type formers and universes in
\cref{sec:universe} presents the transport-heavy intrinsic
syntax problem emphasized by \citet{chen-nordvall-forsberg-tsai>2026}, and its mechanization is
left for future work. Notably, Agda has an
experimental flat modality~\cite{agda>flat,licata-orton-pitts-spitters>2018}, matching the one used in \cref{sec:universe:discrete}. We expect it to be useful for this extension. Agda Synthetic Categories~\cite{toth-bakke>agda-synthetic-categories} develops a broader simplicial type theory library in (non-cubical) Agda. In particular, it formalizes covariant families, localization, and the flat modality, providing infrastructure for many notions used here. This library may be useful for extending the mechanization to universes and parametricity, where more simplicial structure may be needed. Another relevant system is
Rzk~\cite{kudasov-sim-ahrens>2026}, a proof assistant based on simplicial
type theory for synthetic $\infty$-category
theory~\cite{kudasov-riehl-weinberger>2024}. Unlike Rzk, the present mechanization does not develop the
full simplicial layer, such as shapes, but it suffices to formalize directed
gluing and prove canonicity for the fragment with non-dependent type formers.

\subsection{Simplicial Type Theory, Homotopy Type Theory, and Directed Type Theory}
Simplicial type theory is useful here because it combines directed inequalities
and contravariant families with the usual infrastructure of HoTT, including dependent types and HITs. Its main use so
far in the literature has been synthetic $\infty$-category theory
\cite{riehl-shulman>2017,gratzer-weinberger-buchholtz>2024,gratzer-weinberger-buchholtz>2024-yoneda,gratzer-weinberger-buchholtz>2026,weinberger>thesis,weinberger>2024-two-sided,buchholtz-weinberger>2023,martinez>2025,martinez>2025-limits};
the present work uses the same directed structure for the metatheory of syntax
and reduction. Relatedly, \citet{seely>1987} already modeled reduction 2-categorically. An appeal of the present approach is that, because simplicial type theory is a synthetic theory of category theory, we can reason about categorically motivated presentations of reduction in a type-theoretic language.
A future direction to investigate is whether the cocartesian families of \citet{weinberger>thesis} can give
coherent directed coercions for type conversion instead of the discreteness
condition of \cref{sec:universe}.

The present work does not use univalence, nor does it exploit higher structure above the level of sets. Finding applications of
univalence and directed univalence~\cite{weaver-licata>2020,gratzer-weinberger-buchholtz>2024,cavallo-riehl-sattler>2026}
in this setting is an interesting direction for future work.
Throughout the paper, we consider equalities and inequalities between computability witnesses and rarely consider either between computability predicates. In the strict universe $\mathcal{V}$ of \cref{sec:universe}, equality between predicates must be structural. If computability predicates could be constructed in a univalent universe instead, we could consider a coarser (univalent) notion of equality between them, which may help prove representation independence more abstractly \citep{angiuli-cavallo-mortberg-zeuner>2021}.

Other directed type theories often equip inequality types with $\kw{J}$-like
eliminations or more fine-grained variance control
\cite{licata-harper>2011,north>2019,nuyts>2015,neumann-altenkirch>2024,neumann>2025,neumann>thesis,laretto-loregian-voltri>2026}.
These systems can substantially change ordinary type formers, such as
$\Pi$-types, so the present construction may not transfer directly.
In particular, the logic of \citet{laretto-loregian-voltri>2026,laretto-loregian-voltri>2026-arxiv}
can derive a backward transport through syntactic variance control, whereas here
the corresponding operation arises from a semantic proof obligation
that we prove for computability families.
For modeling reductions in directed type theories, \citet{neumann>2025} and
\citet{laretto-loregian-voltri>2026-arxiv} provide the closest points of
contact: both use inequalities to represent reduction, organizing terms
into a preorder. In particular,
\citet[Section~3.1]{laretto-loregian-voltri>2026-arxiv} present a
$\lambda$-calculus signature whose constructors are automatically monotone
with respect to reduction, as in our construction. \Cref{sec:syntax} builds on
these ideas in a slightly different setting.

\subsection{Equational Proof-Relevant Logical Relations}
Existing gluing arguments give proof-relevant logical-relations proofs of
canonicity, normalization, and parametricity for many type theories: simply
typed $\lambda$-theories~\cite{fiore>2002,sterling-spitters>2018}; dependent type
theory~\cite{coquand>2018,altenkirch-kaposi>2017}; abstract canonicity and
parametricity constructions~\cite{kaposi-huber-sattler>2019,bocquet-kaposi-sattler>2023};
homotopy canonicity~\cite{coquand-huber-sattler>2019,shulman>2015}; cubical and
multimodal normalization~\cite{sterling-angiuli>2021,gratzer>2022}; canonicity
with indexed inductive-recursive types~\cite{kovacs>2026}; and proof-relevant
parametricity for ML-style modules~\cite{sterling-harper>2021}. Logical relations over an explicit reduction relation have also been mechanized~\cite{abel-ohman-vezzosi>2017,adjedj-lennon-bertrand-maillard-pedrot-pujet>2024}, where congruence of reduction must be proved; here it is inherited from monotonicity.

These arguments are equational: the syntactic component is typically a CwF or
a signature in a semantic logical framework~\cite{harper>2021}. The logical predicate therefore ranges over
judgmental equivalence classes, so object-language equations are
handled by transport along equalities. Although equality transport always exists, directed transport is available only for
contravariant families. Thus the present work replaces equality by a weaker
directed structure while recovering the needed logical-relations argument.
In \cref{sec:universe}, part of the equational story is recovered by discretizing
syntactic types: judgmentally equal types must induce the same computability
predicate. A more controlled way to recover judgmental equalities may come from
synthetic phase distinctions. Introduced by \citet{sterling-harper>2021} and
used for cost analysis, information flow, and compilation
\cite{niu-sterling-grodin-harper>2022,grodin-niu-sterling-harper>2024,grodin-li-harper>2026,grodin2026potentialfunctionstypes,sterling-harper>2022,theocharis-brady>2026},
phase distinctions add a type-theoretic proposition $\phi$, inducing a two-world Kripke model containing an
``in-the-phase'' world and an ``out-of-the-phase''
world. Judgmental equalities could live in the phase, while
reductions by inequalities live out of it; monotonicity of Kripke worlds is preserved because
reduction is contained in judgmental equality. Synthetically, this requires a
modality that turns inequalities into equalities. In a similar directed setting,
\citet{grodin-niu-sterling-harper>2024} propose a monadic modality $\Op$
satisfying $\Op(x \leq y \to x = y)$, which may serve this purpose. A
proper phase distinction between judgmental equality and reduction is left for
future work, and may be the right way to handle identity types in the directed
setting.

\subsection{Synthetic Tait Computability}
Synthetic Tait computability~\cite{sterling>thesis} gives an abstract
formulation of gluing. It has been used for cubical and modal
type theories~\cite{sterling-angiuli>2021,gratzer>2022},
ML-style modules and effect handlers
\cite{sterling-harper>2021,yang>thesis,yang-wu>2026}, dependent
call-by-push-value~\cite{li-harper>2025}, controlling
unfolding~\cite{gratzer-sterling-angiuli-coquand-birkedal>2025}, and simplicial type theory~\cite{weinberger-ahrens-buchholtz-north>2022}.

The synthetic presentation works in the internal language of the gluing
category. Instead of specifying an external computability predicate
$\stm~\syn{\emp}~(\ssubst{A^\circ}{\gamma^\circ}) \to \Univ$, one specifies an
element of $\Univ$ whose syntactic part is $\stm(A)$, using an extension type
$\{\Univ \mid \kw{syn} \hookrightarrow \stm(A)\}$. This gives a fibered view of
the indexed construction used in the present work. The directed version should
be obtainable by working in the internal language of a suitable gluing model
of simplicial type theory with the required modalities and requiring the
corresponding extension type to be contravariant, recovering the role played
here by the proof-relevant expansion.
A further advantage of this approach is its use of second-order syntax, which
avoids treating contexts and substitutions separately from types and terms and
abstracts away the naturality conditions of CwF-based gluing. This is
particularly useful for normalization~\cite{sterling-angiuli>2021,gratzer>2022,sterling>thesis},
where the logical relation is more involved than in the canonicity proof given
here. A directed version of synthetic Tait computability may help extend the
present method to that setting.

\subsection{Binary Parametricity}
\Cref{sec:binary} extends the construction to proof-relevant parametricity, an analytic reconstruction of the synthetic left and right syntactic modalities of \citet{sterling-harper>2021}.
The key point is to separate homogeneous vertical relations for judgmental
equalities or reductions ($M_L$ \vs $M'_L$ and $M_R$ \vs $M'_R$) from
heterogeneous horizontal relations for parametricity ($M_L$ \vs $M_R$ and
$M'_L$ \vs $M'_R$).
\[
\centering
\begin{tikzcd}[column sep=5.5em,row sep=large,labels={font=\footnotesize\everymath{\textstyle}}]
  M_L
    \arrow[d,draw=oc-red-9,"{r_L : M_L \leq M'_L}"']
    \arrow[r,draw=oc-blue-9,dashed,no head,
      "{\Phi : A^\bullet~M_L~M_R}"]
    &
  M_R
    \arrow[d,draw=oc-red-9,"{r_R : M_R \leq M'_R}"]
  \\
  M'_L
    \arrow[r,draw=oc-blue-9,dashed,no head,
      "{\Phi' : A^\bullet~M'_L~M'_R}"']
    &
  M'_R
\end{tikzcd}
\]
In an impredicative setting, one would use a quasi-PER binary relation to account for both judgmental equality and
parametricity~\cite{krishnaswami-dreyer>2013}. The closure condition for quasi-PERs in that setting amounts to
moving along the vertical homogeneous dimension inside the horizontal heterogeneous
relation here. Categorically, the parametricity relation here, as in
\citet{sterling-harper>2021}, is a span
$\stm~\syn{\emp}~A_L^\circ \xleftarrow{s} A^\bullet \xrightarrow{t}
\stm~\syn{\emp}~A_R^\circ$, writing $A^\bullet$ for the total space of the relation; an element $M^\bullet : A^\bullet$ witnesses that
$M_L = s(M^\bullet)$ and $M_R = t(M^\bullet)$ are parametrically related.
A quasi-PER can be viewed
as a span equipped with heterogeneous transitivity. The vertical/horizontal
separation clarifies that contravariance belongs to the vertical
dimension: it transports computability along reductions, rather than expressing
parametricity itself. It also suggests a double-categorical semantics for
parametricity, which we leave for future work.
 
\section*{Data Availability Statement}

The fragment of directed logical relations with non-dependent type formers, described in \cref{sec:syntax,sec:gluing}, is mechanized in Cubical Agda~\citep{norell>2009,vezzosi-mortberg-abel>2019}. The directed syntax is constructed via higher inductive types and reflective subuniverses as in \cref{sec:syntax:initial-directed-cwf}. The logical relation covers products, functions, and Booleans, with a fundamental theorem that implies directed Boolean canonicity. The accompanying mechanization~\citep{li-grodin-harper-artifact>2026} provides a \textsf{README.agda} that maps Agda definitions to the definitions, theorems, and constructions presented in this paper. A browsable HTML rendering of the mechanization is available at
\begin{center}
  \url{https://www.cs.cmu.edu/~runmingl/dprlr/DPRLR.README.html}.
\end{center}

\section*{Disclosure of Use of AI in Research}
The accompanying artifact was written with the assistance of an AI agent. In particular, we found the AI tool effective in
generating repetitive code patterns, of which there are many in a formalization of metatheory.
The authors made all design decisions, including what to prove and how to prove it. AI was used only to help
fill in some implementation details and generate repetitive code for proofs whose methods had already been determined by the
authors. The authors take full responsibility for the entire codebase. AI was not used outside the artifact.

\begin{acks}
  The authors thank Jonathan Sterling and Yue Yao for adjacent collaborations and Daniel Gratzer and Ulrik Buchholtz for suggesting the use of the flat modality in simplicial type theory.

  This material is based upon work supported by the \grantsponsor{AFOSR}{United States Air Force Office of Scientific Research}{https://www.afrl.af.mil/AFOSR/} under grant numbers \grantnum{AFOSR}{FA9550-21-0009} and \grantnum{AFOSR}{FA9550-23-1-0434} (Tristan Nguyen, program manager) and the \grantsponsor{NSF}{National Science Foundation}{https://nsf.gov} under award number \grantnum{NSF}{2615896}.
  Any opinions, findings and conclusions or recommendations expressed in this material are those of the authors and do not necessarily reflect the views of AFOSR or NSF.
  The authors are grateful for support from \grantsponsor{JSC}{Jane Street Capital}{https://www.janestreet.com/}.
\end{acks}

\paperreferences
\fi

\iftoggle{arxiv}{
\ifappendixonly\else\clearpage\fi
\appendix
\raggedbottom
\section{Orthogonality Constructions}\label[appendix]{appendix:preorders}

A standard way to construct a reflective subuniverse in HoTT is by
localization at a family of maps~\cite{christensen-opie-rijke-scoccola>2020,rijke-shulman-spitters>2020}. The resulting subuniverse consists of
types that are local at each map in the family, in the following sense.

\begin{definition}[Orthogonality]
For a map $f:Y\to Z$, a type $X$ is \emph{orthogonal} to $f$, or
\emph{$f$-local}, when precomposition with $f$ is an equivalence:
\[
  (-\circ f) : (Z\to X) \to (Y\to X).
\]
For a family of maps $\mathcal{F}$, write $\kw{isLocal}_{\mathcal{F}}~X$
when $X$ is local for every map in $\mathcal{F}$.
\end{definition}

Many familiar properties of types can be expressed as orthogonality
conditions. The three conditions needed in \maincref{sec:reflective-subuniverse} to construct synthetic preorders are:
\begin{enumerate}[wide=4pt]
\item \emph{Setness.} A type is a set exactly when it is local at the
  terminal map $f_{\kw{set}}\isdef~ !:S^1\to\mathbf{1}$ from the circle
  \cite[Example~2.3(v)]{rijke-shulman-spitters>2020}. A map from $S^1$ to
  $X$ consists of a point $x:X$ and a loop $p:x=x$. Precomposition with
  $!$ assigns the constant loop to each point, and is an equivalence
  exactly when each loop space is contractible, equivalently when $X$ is a set.

\item \emph{Thinness.} Following
  \citet[Definition~4.12]{grodin-niu-sterling-harper>2024}, define the
  \emph{path suspension} $\PathSusp(Y)$ as the following pushout. Its
  universal property, displayed on the right, says that a map into $X$
  consists of two points and a $Y$-indexed family of inequalities between
  them.
\par\noindent
\begin{minipage}[c]{.43\linewidth}
\centering
\begin{tikzcd}[ampersand replacement=\&, row sep=1.7em, column sep=4.5em]
  Y\times\{0,1\}
    \arrow[r, "{\kw{id}\times[\mathsf{i0}\mid\mathsf{i1}]}"]
    \arrow[d, "{\pi_2}"']
  \& Y\times\mathbbm{2} \arrow[d] \\
  \{0,1\} \arrow[r]
  \& \PathSusp(Y) \arrow[ul, phantom, very near start, "\ulcorner"]
\end{tikzcd}
\end{minipage}\hfill
\begin{minipage}[c]{.53\linewidth}
\[
  (\PathSusp(Y)\to X)
  \simeq
  \sum_{x,x':X}(Y\to x\leq_X x').
\]
\end{minipage}
\par
  For $!: \{0,1\}\to\mathbf{1}$, locality at the induced map
  $f_{\kw{thin}}\isdef\PathSusp(!):
    \PathSusp(\{0,1\})\to\PathSusp(\mathbf{1})$ says that the map $(x\leq_X x')\to(\{0,1\}\to x\leq_X x')$ is an
  equivalence. This holds exactly when each inequality type is a
  proposition, so locality at $f_{\kw{thin}}$ is thinness
  \cite[Lemma~4.13]{grodin-niu-sterling-harper>2024}.

\item \emph{Segalness.}
  In simplicial type theory, the ``transitive'' property of inequality at a type is captured by the \emph{Segal} condition. Let $\Delta^2$ and $\Lambda^2_1$ be the following shapes (where $\leq_{\mathbbm{2}}$ is the bounded order on the interval):
  \[
    \Delta^2 \isdef
    \{(s,t) : \mathbbm{2} \times \mathbbm{2} \mid t \leq_{\mathbbm{2}} s\},
    \qquad
    \Lambda^2_1 \isdef
    \{(s,t) : \mathbbm{2} \times \mathbbm{2}
      \mid s = \mathsf{i1} \lor t = \mathsf{i0}\}.
  \]
  Write $\iota : \Lambda^2_1 \hookrightarrow \Delta^2$ for the inclusion.
  Intuitively, a map $\Lambda^2_1\to X$ consists of two directed paths
  $p:x\leq_X x'$ and $q:x'\leq_X x''$, without a chosen composite. An extension
  to $\Delta^2\to X$ supplies a path $r:x\leq_X x''$ and a 2-simplex witnessing
  $r$ as a composite of $p$ and $q$. Such an extension need not exist for an
  arbitrary type $X$.
  A type $X$ is \emph{Segal} when restriction along $\iota$ is an equivalence
  \cite[Theorem~5.5]{riehl-shulman>2017}:
  \[
    \kw{isSegal}~X
    \isdef
    \kw{isEquiv}~\bigl((-\circ\iota) :
      (\Delta^2 \to X) \to (\Lambda^2_1 \to X)\bigr).
  \]
  Thus Segalness is locality at
  $f_{\kw{Segal}}\isdef\iota:\Lambda^2_1\hookrightarrow\Delta^2$.
\end{enumerate}

Write $\mathcal{F}_p\isdef\{f_{\kw{set}},f_{\kw{thin}},f_{\kw{Segal}}\}$.
The promised subuniverse is
\[
  \mathcal{U}_p
  \isdef
  \sum_{X:\Univ}\kw{isLocal}_{\mathcal{F}_p}~X.
\]

The inclusion $\mathcal{U}_p \hookrightarrow \Univ$ of the subuniverse of preorders into the ambient universe comes equipped with a left adjoint \emph{reflector} $\mathsf{P} : \Univ \to \mathcal{U}_p$ and a modal unit $\eta_X:X\to\mathsf{P} X$.

\begin{lemma}[Universal property of $\mathsf{P}$; {\protect\citealp[Lem.~2.17]{rijke-shulman-spitters>2020}} \AgdaFormalized]
\label{lem:localization-universal}
For every $X:\Univ$ and $Y:\mathcal{U}_p$,
precomposition with $\eta_X$ is an equivalence:
\[
  (-\circ\eta_X): (\mathsf{P} X\to Y) \simeq (X\to Y).
\]
\end{lemma}

\subsection{Discrete Preorders}\label[appendix]{appendix:preorders:discrete}

Discreteness is also an orthogonality condition, so the localization
construction applies.
\begin{lemma}[Discreteness as orthogonality; {\protect\citealp[Def.~2.10]{gratzer-weinberger-buchholtz>2024}}]\label{lem:discrete-orthogonal}
  A type $X$ is discrete if and only if it is local at the terminal map
  $f_{\kw{disc}}\isdef ~!:\mathbbm{2}\to\mathbf{1}$.
\end{lemma}
Together with the setness map $f_{\kw{set}}$, this gives the reflective subuniverse
of discrete sets:
\[
  \mathcal{F}_s\isdef\{f_{\kw{set}},f_{\kw{disc}}\},
  \qquad
  \mathcal{U}_s\isdef
    \sum_{X:\Univ}\kw{isLocal}_{\mathcal{F}_s}~X.
\]
Discrete sets are preorders, so $\mathcal{U}_s$ is a subuniverse of
$\mathcal{U}_p$.

\clearpage
\section{Reference Definitions}\label[appendix]{sec:reference-definitions}

This appendix is a reference sheet for the final glued structures and the (non-dependent and dependent) type formers.  The construction uses the flat predicates of
\maincref{sec:universe} and should be read as the flat refinement of the
construction first presented in \maincref{sec:gluing}. Predicates take values
in the strict universe $\mathcal{V}$ of \maincref{sec:universe} and universe levels
and decoding are left implicit.

We do not repeat the $\beta$- and $\eta$-laws for the type formers.  In each
case, the proof has the same shape: contravariant transport of the semantic
witness along the relevant syntactic reduction is compared with the original
witness, and the universal property of contravariance
(\maincref{lem:contrav-universal}) turns this comparison into the same reflexivity
argument used in \maincref{sec:gluing}.  We write $\kappa$ for the inverse of the
counit $\eps$ on flat modal types, following \maincref{def:flat-modal}.

\subsection{Unary Model}

\subsubsection{Flat Predicates}

For a closed syntactic type $A^\circ : \sty~\syn{\emp}$, the flat predicate is:
\begin{logrel}
\begin{array}{@{}l@{\;}c@{\;}l@{}}
  \Pred(A^\circ)
  &\isdef&
  \kw{let}^{\flat}~B = \kappa(A^\circ)~\kw{in}~
  \flat(\stm~\syn{\emp}~B \to \mathcal{V}).
\end{array}
\end{logrel}

\subsubsection{Judgmental Structure}

The final unary judgmental structures are:
\begin{logrel}
\begin{array}{@{\guiderule{oc-blue-9}}l@{}}
  \declkw{record}~\glu{\kw{CTX}} : \mathcal{U}_s~\declkw{where} \\
  \quad \Gamma^\circ : \sctx \\
  \quad \Gamma^\bullet_\flat :
    \flat(\ssub~\syn{\emp}~\Gamma^\circ \to \mathcal{V}) \\
  \quad \Gamma^\bullet \isdef \eps(\Gamma^\bullet_\flat) \\
  \quad c_\Gamma : \kw{isContrav}~\Gamma^\bullet
  \\ \noalign{\vskip 8pt}
  \declkw{record}~\glu{\kw{SUB}}~
    (\Gamma~\Delta : \glu{\kw{CTX}}) : \mathcal{U}_p~\declkw{where} \\
  \quad \sigma^\circ :
    \ssub~\Gamma^\circ~\Delta^\circ \\
  \quad \sigma^\bullet :
    (\gamma^\circ : \ssub~\syn{\emp}~\Gamma^\circ) \to
    \Gamma^\bullet~\gamma^\circ \to
    \Delta^\bullet~(\sigma^\circ \scomp \gamma^\circ)
  \\ \noalign{\vskip 8pt}
  \declkw{record}~\glu{\kw{TY}}~
    (\Gamma : \glu{\kw{CTX}}) : \mathcal{U}_s~\declkw{where} \\
  \quad A^\circ : \sty~\Gamma^\circ \\
  \quad A^\bullet_\flat :
    (\gamma^\circ : \ssub~\syn{\emp}~\Gamma^\circ) \to
    (\gamma^\bullet : \Gamma^\bullet~\gamma^\circ) \to
    \Pred(\ssubst{A^\circ}{\gamma^\circ}) \\
  \quad A^\bullet~\gamma^\circ~\gamma^\bullet
    \isdef \eps(A^\bullet_\flat~\gamma^\circ~\gamma^\bullet) \\
  \quad c_A :
    (\gamma^\circ : \ssub~\syn{\emp}~\Gamma^\circ) \to
    (\gamma^\bullet : \Gamma^\bullet~\gamma^\circ) \to
    \kw{isContrav}~(A^\bullet~\gamma^\circ~\gamma^\bullet)
  \\ \noalign{\vskip 8pt}
  \declkw{record}~\glu{\kw{TM}}~
    (\Gamma : \glu{\kw{CTX}})~
    (A : \glu{\kw{TY}}~\Gamma) : \mathcal{U}_p~\declkw{where} \\
  \quad M^\circ : \stm~\Gamma^\circ~A^\circ \\
  \quad M^\bullet :
    (\gamma^\circ : \ssub~\syn{\emp}~\Gamma^\circ) \to
    (\gamma^\bullet : \Gamma^\bullet~\gamma^\circ) \to
    A^\bullet~\gamma^\circ~\gamma^\bullet~
      (\ssubst{M^\circ}{\gamma^\circ})
\end{array}
\end{logrel}

\begin{samepage}
\subsubsection{Substitutions and Context Extension}

The substitution definitions are:
\begin{logrel}
\begin{array}{@{}l@{\;}c@{\;}l@{}}
  \glu{\kw{ID}}^\circ
    &\isdef& \sidt
  \\
  \glu{\kw{ID}}^\bullet~\gamma^\circ~\gamma^\bullet
    &\isdef& \gamma^\bullet
  \\[8pt]
  (\tau \gcomp \sigma)^\circ
    &\isdef& \tau^\circ \scomp \sigma^\circ
  \\
  (\tau \gcomp \sigma)^\bullet~\gamma^\circ~\gamma^\bullet
    &\isdef&
    \tau^\bullet~
      (\sigma^\circ \scomp \gamma^\circ)~
      (\sigma^\bullet~\gamma^\circ~\gamma^\bullet)
  \\[8pt]
  (A\glu{[}\sigma\glu{]})^\circ
    &\isdef& \ssubst{A^\circ}{\sigma^\circ}
  \\
  (A\glu{[}\sigma\glu{]})^\bullet_\flat~\gamma^\circ~\gamma^\bullet
    &\isdef&
    A^\bullet_\flat~
      (\sigma^\circ \scomp \gamma^\circ)~
      (\sigma^\bullet~\gamma^\circ~\gamma^\bullet)
  \\
  c_{A\glu{[}\sigma\glu{]}}~\gamma^\circ~\gamma^\bullet
    &\isdef&
    c_A~
      (\sigma^\circ \scomp \gamma^\circ)~
      (\sigma^\bullet~\gamma^\circ~\gamma^\bullet)
  \\[8pt]
  (M\glu{[}\sigma\glu{]})^\circ
    &\isdef& \ssubst{M^\circ}{\sigma^\circ}
  \\
  (M\glu{[}\sigma\glu{]})^\bullet~\gamma^\circ~\gamma^\bullet
    &\isdef&
    M^\bullet~
      (\sigma^\circ \scomp \gamma^\circ)~
      (\sigma^\bullet~\gamma^\circ~\gamma^\bullet)
\end{array}
\end{logrel}
\end{samepage}
\begin{logrel}
\begin{array}{@{}l@{}}
  \glu{\emp}^\circ
    \isdef \syn{\emp}
  \\[3pt]
  \glu{\emp}^\bullet_\flat
    \isdef
    \bigl(\lambda \gamma^\circ.\; \unit\bigr)^\flat
  \\[8pt]
  (\Gamma \gext A)^\circ
    \isdef \Gamma^\circ \sext A^\circ
  \\[3pt]
  (\Gamma \gext A)^\bullet_\flat
    \isdef
    \kw{let}^{\flat}~\Gamma^\bullet = \Gamma^\bullet_\flat~\kw{in}~
    \kw{let}^{\flat}~\mathcal{A} =
      \kappa(A^\bullet_\flat)~\kw{in}~\\
    \quad
    \left(
      \begin{array}{@{}l@{}}
        \lambda \delta^\circ.\;
        \displaystyle
        \sum_{\delta^\bullet :
          \Gamma^\bullet~(\spp \scomp \delta^\circ)}
        \kw{let}^{\flat}~A^\bullet =
          \mathcal{A}~(\spp \scomp \delta^\circ)~\delta^\bullet~\kw{in}~
        A^\bullet~(\ssubst{\sqq}{\delta^\circ})
      \end{array}
    \right)^\flat
  \\[8pt]
  \glu{(}\sigma\glu{,} M\glu{)}^\circ
    \isdef \spair{\sigma^\circ}{M^\circ}
  \\
  \glu{(}\sigma\glu{,} M\glu{)}^\bullet~\gamma^\circ~\gamma^\bullet
    \isdef
    \bigl(\sigma^\bullet~\gamma^\circ~\gamma^\bullet,\,
      M^\bullet~\gamma^\circ~\gamma^\bullet\bigr)
  \\[8pt]
  (\glu{\kw{p}})^\circ
    \isdef \spp
  \\
  (\glu{\kw{p}})^\bullet~\delta^\circ~(\delta^\bullet,a^\bullet)
    \isdef \delta^\bullet
  \\[8pt]
  (\glu{\kw{q}})^\circ
    \isdef \sqq
  \\
  (\glu{\kw{q}})^\bullet~\delta^\circ~(\delta^\bullet,a^\bullet)
    \isdef a^\bullet
\end{array}
\end{logrel}
Taking the counit of the context-extension predicate recovers the ordinary
non-flat predicate:
\begin{logrel}
\begin{array}{@{}l@{}}
  \eps((\Gamma \gext A)^\bullet_\flat)~\delta^\circ
  =
  \displaystyle
  \sum_{\delta^\bullet :
    \eps(\Gamma^\bullet_\flat)~(\spp \scomp \delta^\circ)}
  \eps\bigl(
    A^\bullet_\flat~
    (\spp \scomp \delta^\circ)~
    \delta^\bullet
  \bigr)~
  (\ssubst{\sqq}{\delta^\circ}) .
\end{array}
\end{logrel}

The contravariance of the empty context and context extension predicates is established by the contravariance of the constant unit predicate and the closure of contravariant predicates under dependent sums (\cref{lem:contrav-sigma}), respectively.

\subsubsection{Non-Dependent Products}

The non-dependent product type definitions are:
\begin{logrel}
\begin{array}{@{}l@{}}
  \glu{\kw{PROD}} :
    \glu{\kw{TY}}~\Gamma \to
    \glu{\kw{TY}}~\Gamma \to
    \glu{\kw{TY}}~\Gamma
  \\[3pt]
  (\glu{\kw{PROD}}~A~B)^\circ
    \isdef A^\circ \mathbin{\syn{\times}} B^\circ
  \\[3pt]
  (\glu{\kw{PROD}}~A~B)^\bullet_\flat~\gamma^\circ~\gamma^\bullet
    \isdef
    \kw{let}^{\flat}~A^\bullet = A^\bullet_\flat~\gamma^\circ~\gamma^\bullet~\kw{in}~
    \kw{let}^{\flat}~B^\bullet = B^\bullet_\flat~\gamma^\circ~\gamma^\bullet~\kw{in}\\
    \quad
    \bigl(
      \lambda P.\;
        A^\bullet~(\syn{\kw{fst}}~P)
        \times
        B^\bullet~(\syn{\kw{snd}}~P)
    \bigr)^\flat
\end{array}
\end{logrel}
For $c_{\glu{\kw{PROD}}~A~B}$, reindex $c_A$ and $c_B$ along
$\syn{\kw{fst}}$ and $\syn{\kw{snd}}$, then apply
\maincref{lem:contrav-reindex,lem:contrav-product}.

\begin{logrel}
\begin{array}{@{}l@{}}
  \glu{\kw{PAIR}} :
    \glu{\kw{TM}}~\Gamma~A \to
    \glu{\kw{TM}}~\Gamma~B \to
    \glu{\kw{TM}}~\Gamma~(\glu{\kw{PROD}}~A~B)
  \\[3pt]
  (\glu{\kw{PAIR}}~M~N)^\circ
    \isdef \syn{\kw{pair}}~M^\circ~N^\circ
  \\
  (\glu{\kw{PAIR}}~M~N)^\bullet~
    \gamma^\circ~\gamma^\bullet
    \isdef
    \Bigl(
      (\ssubst{\syn{\times_{\beta_1}}}{\gamma^\circ})^*~
        (M^\bullet~\gamma^\circ~\gamma^\bullet),\,
      (\ssubst{\syn{\times_{\beta_2}}}{\gamma^\circ})^*~
        (N^\bullet~\gamma^\circ~\gamma^\bullet)
    \Bigr)
  \\[8pt]
  (\glu{\kw{FST}}~P)^\circ
    \isdef \syn{\kw{fst}}~P^\circ
  \\[3pt]
  (\glu{\kw{FST}}~P)^\bullet~
    \gamma^\circ~\gamma^\bullet
    \isdef \pi_1(P^\bullet~\gamma^\circ~\gamma^\bullet)
  \\[8pt]
  (\glu{\kw{SND}}~P)^\circ
    \isdef \syn{\kw{snd}}~P^\circ
  \\[3pt]
  (\glu{\kw{SND}}~P)^\bullet~
    \gamma^\circ~\gamma^\bullet
    \isdef \pi_2(P^\bullet~\gamma^\circ~\gamma^\bullet)
\end{array}
\end{logrel}

\subsubsection{Non-Dependent Functions}

The non-dependent function type definitions are:
\begin{logrel}
\begin{array}{@{}l@{}}
  \glu{\kw{FUN}} :
    \glu{\kw{TY}}~\Gamma \to
    \glu{\kw{TY}}~\Gamma \to
    \glu{\kw{TY}}~\Gamma
  \\[3pt]
  (\glu{\kw{FUN}}~A~B)^\circ
    \isdef A^\circ \mathbin{\syn{\Rightarrow}} B^\circ
  \\[3pt]
  (\glu{\kw{FUN}}~A~B)^\bullet_\flat~
    \gamma^\circ~\gamma^\bullet
    \isdef
    \kw{let}^{\flat}~A^\bullet = A^\bullet_\flat~\gamma^\circ~\gamma^\bullet~\kw{in}~
    \kw{let}^{\flat}~B^\bullet = B^\bullet_\flat~\gamma^\circ~\gamma^\bullet~\kw{in}\\
    \quad
    \left(
      \begin{array}{@{}l@{}}
        \lambda F.\;
          (M^\circ : \stm~\syn{\emp}~(\ssubst{A^\circ}{\gamma^\circ})) \to \\
        \qquad
          A^\bullet~M^\circ \to
          B^\bullet~(\syn{\kw{app}}~F~M^\circ)
      \end{array}
    \right)^\flat
\end{array}
\end{logrel}
For $c_{\glu{\kw{FUN}}~A~B}$, reindex $c_B$ along application at each
fixed closed argument and witness, then apply
\maincref{lem:contrav-reindex,lem:contrav-pi}, as in
\maincref{sec:canonicity:function}.

\begin{logrel}
\begin{array}{@{}l@{}}
  \glu{\kw{LAM}} :
    \glu{\kw{TM}}~(\Gamma \gext A)~
      (B\glu{[}\glu{\kw{p}}\glu{]}) \to
    \glu{\kw{TM}}~\Gamma~(\glu{\kw{FUN}}~A~B)
  \\[3pt]
  (\glu{\kw{LAM}}~N)^\circ
    \isdef \syn{\kw{lam}}~N^\circ
  \\
  (\glu{\kw{LAM}}~N)^\bullet~
    \gamma^\circ~\gamma^\bullet~M^\circ~M^\bullet
    \isdef
    (\ssubst{\syn{\Rightarrow_{\beta}}}
      {\spair{\gamma^\circ}{M^\circ}})^*
    \bigl(
      N^\bullet~
        \spair{\gamma^\circ}{M^\circ}~
        (\gamma^\bullet,M^\bullet)
    \bigr)
  \\[8pt]
  \glu{\kw{APP}} :
    \glu{\kw{TM}}~\Gamma~(\glu{\kw{FUN}}~A~B) \to
    \glu{\kw{TM}}~\Gamma~A \to
    \glu{\kw{TM}}~\Gamma~B
  \\[3pt]
  (\glu{\kw{APP}}~F~M)^\circ
    \isdef \syn{\kw{app}}~F^\circ~M^\circ
  \\
  (\glu{\kw{APP}}~F~M)^\bullet~
    \gamma^\circ~\gamma^\bullet
    \isdef
    F^\bullet~\gamma^\circ~\gamma^\bullet~
      (\ssubst{M^\circ}{\gamma^\circ})~
      (M^\bullet~\gamma^\circ~\gamma^\bullet)
\end{array}
\end{logrel}

\subsubsection{Revisiting Dependency}\label{sec:universe:dependency-revisited}

For $A:\glu{\kw{TY}}~\Gamma$, take an inequality of closed glued substitutions
\[
  (\rho^\circ,\rho^\bullet):
    (\gamma_0^\circ,\gamma_0^\bullet)
    \leq(\gamma_1^\circ,\gamma_1^\bullet).
\]
By monotonicity, the predicates are related over the induced type
reduction $r\isdef \ssubst{A^\circ}{\rho^\circ}$:
\[
  A^\bullet_\flat~\gamma_0^\circ~\gamma_0^\bullet
  \leq_{\Pred(\ssubst{A^\circ}{\rho^\circ})}
  A^\bullet_\flat~\gamma_1^\circ~\gamma_1^\bullet.
\]
Discreteness of syntactic types turns $r$ into a path
$e_r:\ssubst{A^\circ}{\gamma_0^\circ}
     =\ssubst{A^\circ}{\gamma_1^\circ}$.
Each fiber of $\Pred$ is flat modal, hence discrete.
Thus the dependent inequality induces an equality after transport along $e_r$:
\[
  \kw{transport}_{\Pred}(e_r)
    (A^\bullet_\flat~\gamma_0^\circ~\gamma_0^\bullet)
  = A^\bullet_\flat~\gamma_1^\circ~\gamma_1^\bullet.
\]
Applying the counit gives an equality of computability predicates.
Transport in the predicate domain becomes precomposition with inverse
conversion:
\[
\begin{aligned}
  A^\bullet~\gamma_0^\circ~\gamma_0^\bullet\circ\kw{conv}_r^{-1}
    &=\eps\bigl(\kw{transport}_{\Pred}(e_r)
        (A^\bullet_\flat~\gamma_0^\circ~\gamma_0^\bullet)\bigr)\\
    &=\eps(A^\bullet_\flat~\gamma_1^\circ~\gamma_1^\bullet)\\
    &=A^\bullet~\gamma_1^\circ~\gamma_1^\bullet.
\end{aligned}
\]
Evaluating at
$M_1^\circ:\stm~\syn{\emp}~(\ssubst{A^\circ}{\gamma_1^\circ})$
gives the substitution-invariance path:
\[
\begin{aligned}
  \kw{substInv}_{A,\rho,M_1^\circ}:\quad
  &A^\bullet~\gamma_0^\circ~\gamma_0^\bullet~
    (\kw{conv}_r^{-1}~M_1^\circ)\\
  &\qquad = A^\bullet~\gamma_1^\circ~\gamma_1^\bullet~M_1^\circ.
\end{aligned}
\]
Intuitively, reduction-related glued substitutions into a syntactic type
induce equal computability predicates, up to the induced type conversion.

\subsubsection{Dependent Pairs}
The dependent pair type definitions are:
\begin{logrel}
\begin{array}{@{}l@{}}
  \glu{\Sigma} :
    (A : \glu{\kw{TY}}~\Gamma) \to
    \glu{\kw{TY}}~(\Gamma \gext A) \to
    \glu{\kw{TY}}~\Gamma
  \\[3pt]
  (\glu{\Sigma}~A~B)^\circ
    \isdef \syn{\Sigma}~A^\circ~B^\circ
  \\[3pt]
  (\glu{\Sigma}~A~B)^\bullet_\flat~
    \gamma^\circ~\gamma^\bullet
    \isdef
    \kw{let}^{\flat}~A^\bullet = A^\bullet_\flat~\gamma^\circ~\gamma^\bullet~\kw{in}~
    \kw{let}^{\flat}~\mathcal{B} =
      \kappa\left(
        \begin{array}{@{}l@{}}
          \lambda M^\circ~M^\bullet.\;
            B^\bullet_\flat~
              \spair{\gamma^\circ}{M^\circ}~
              (\gamma^\bullet,M^\bullet)
        \end{array}
      \right)~\kw{in}~\\
    \quad
    \left(
      \begin{array}{@{}l@{}}
        \lambda P.\;
        \displaystyle
        \sum_{\Phi :
          A^\bullet~(\syn{\kw{fst}}~P)}
        \kw{let}^{\flat}~B^\bullet = \mathcal{B}~(\syn{\kw{fst}}~P)~\Phi~\kw{in}~
        B^\bullet~(\syn{\kw{snd}}~P)
      \end{array}
    \right)^\flat
\end{array}
\end{logrel}
\begin{lemma}[Contravariance under $\Sigma$-types \AgdaFormalized]\label{lem:contrav-sigma}
  Let $Y:X\to\Univ$ be contravariant.
  If $Z:(\sum_{x:X}Y(x))\to\Univ$ is also contravariant, then so is
  \[
    x \mapsto \sum_{y:Y(x)}Z(x,y).
  \]
\end{lemma}

The proof $c_{\glu{\Sigma}~A~B}$ follows from \cref{lem:contrav-sigma},
using $c_A$ for the first-component witnesses and $c_B$, together with
flatness of predicates, for the second-component witnesses over
their total space.

\begin{logrel}
\begin{array}{@{}l@{}}
  \glu{\kw{PAIR}} :
    (M : \glu{\kw{TM}}~\Gamma~A) \to
    \glu{\kw{TM}}~\Gamma~
      (B\glu{[}\glu{(}\glu{\kw{ID}}\glu{,}M\glu{)}\glu{]}) \to
    \glu{\kw{TM}}~\Gamma~(\glu{\Sigma}~A~B)
  \\[3pt]
  (\glu{\kw{PAIR}}~M~N)^\circ
    \isdef
    \syn{\kw{pair}}~M^\circ~N^\circ
  \\
  (\glu{\kw{PAIR}}~M~N)^\bullet~
    \gamma^\circ~\gamma^\bullet
    \isdef
    \left(
      \begin{array}{@{}l@{}}
        (\ssubst{\syn{\Sigma_{\beta_1}}}{\gamma^\circ})^*
          (M^\bullet~\gamma^\circ~\gamma^\bullet),
        \\[3pt]
        (\ssubst{\syn{\Sigma_{\beta_2}}}{\gamma^\circ})^*
          \bigl(\kw{transport}(\kw{substInv}^{-1})
            (N^\bullet~\gamma^\circ~\gamma^\bullet)\bigr)
      \end{array}
    \right)
\end{array}
\end{logrel}
This typechecks as follows, writing
$\Phi\isdef(\ssubst{\syn{\Sigma_{\beta_1}}}{\gamma^\circ})^*
  (M^\bullet~\gamma^\circ~\gamma^\bullet)$:
{\small
\[
\begin{aligned}
  N^\bullet~\gamma^\circ~\gamma^\bullet
  & : B^\bullet~\spair{\gamma^\circ}{\ssubst{M^\circ}{\gamma^\circ}}~
      (\gamma^\bullet,M^\bullet~\gamma^\circ~\gamma^\bullet)~
        (\ssubst{N^\circ}{\gamma^\circ}),\\[3pt]
  \kw{transport}(\kw{substInv}^{-1})
      (N^\bullet~\gamma^\circ~\gamma^\bullet)
  & : B^\bullet~\spair{\gamma^\circ}{\syn{\kw{fst}}~
        \ssubst{(\syn{\kw{pair}}~M^\circ~N^\circ)}{\gamma^\circ}}~
      (\gamma^\bullet,\Phi)~
        \bigl(\kw{conv}^{-1}~\ssubst{N^\circ}{\gamma^\circ}\bigr),\\[3pt]
  (\ssubst{\syn{\Sigma_{\beta_2}}}{\gamma^\circ})^*
      \bigl(\kw{transport}(\kw{substInv}^{-1})
        (N^\bullet~\gamma^\circ~\gamma^\bullet)\bigr)
  & : B^\bullet~\spair{\gamma^\circ}{\syn{\kw{fst}}~
        \ssubst{(\syn{\kw{pair}}~M^\circ~N^\circ)}{\gamma^\circ}}~
      (\gamma^\bullet,\Phi) \\
  & \qquad \bigl(\syn{\kw{snd}}~
        \ssubst{(\syn{\kw{pair}}~M^\circ~N^\circ)}{\gamma^\circ}\bigr).
\end{aligned}
\]
}

\begin{logrel}
\begin{array}{@{}l@{}}
  \glu{\kw{FST}} :
    \glu{\kw{TM}}~\Gamma~(\glu{\Sigma}~A~B) \to
    \glu{\kw{TM}}~\Gamma~A
  \\
  (\glu{\kw{FST}}~P)^\circ
    \isdef \syn{\kw{fst}}~P^\circ
  \\[3pt]
  (\glu{\kw{FST}}~P)^\bullet~
    \gamma^\circ~\gamma^\bullet
    \isdef \pi_1(P^\bullet~\gamma^\circ~\gamma^\bullet)
  \\[8pt]
  \glu{\kw{SND}} :
    (P : \glu{\kw{TM}}~\Gamma~(\glu{\Sigma}~A~B)) \to
    \glu{\kw{TM}}~\Gamma~
      (B\glu{[}\glu{(}\glu{\kw{ID}}\glu{,}\glu{\kw{FST}}~P\glu{)}\glu{]})
  \\
  (\glu{\kw{SND}}~P)^\circ
    \isdef \syn{\kw{snd}}~P^\circ
  \\[3pt]
  (\glu{\kw{SND}}~P)^\bullet~
    \gamma^\circ~\gamma^\bullet
    \isdef \pi_2(P^\bullet~\gamma^\circ~\gamma^\bullet)
\end{array}
\end{logrel}

\subsubsection{Dependent Functions}

The dependent function type definitions are:
\begin{logrel}
\begin{array}{@{}l@{}}
  \glu{\Pi} :
    (A : \glu{\kw{TY}}~\Gamma) \to
    \glu{\kw{TY}}~(\Gamma \gext A) \to
    \glu{\kw{TY}}~\Gamma
  \\[3pt]
  (\glu{\Pi}~A~B)^\circ
    \isdef \syn{\Pi}~A^\circ~B^\circ
  \\[3pt]
  (\glu{\Pi}~A~B)^\bullet_\flat~
    \gamma^\circ~\gamma^\bullet
    \isdef
    \kw{let}^{\flat}~A^\bullet = A^\bullet_\flat~\gamma^\circ~\gamma^\bullet~\kw{in}~
    \kw{let}^{\flat}~\mathcal{B} =
      \kappa\left(
        \begin{array}{@{}l@{}}
          \lambda M^\circ~M^\bullet.\;
            B^\bullet_\flat~
              \spair{\gamma^\circ}{M^\circ}~
              (\gamma^\bullet,M^\bullet)
        \end{array}
      \right)~\kw{in}~\\
      \quad
    \left(
      \begin{array}{@{}l@{}}
        \lambda F.\;
        (M^\circ :
          \stm~\syn{\emp}~
            (\ssubst{A^\circ}{\gamma^\circ}))
        \to
        \\[3pt]
        \qquad
        (M^\bullet : A^\bullet~M^\circ) \to
        \kw{let}^{\flat}~B^\bullet = \mathcal{B}~M^\circ~M^\bullet~\kw{in}~
        B^\bullet~(\syn{\kw{app}}~F~M^\circ)
      \end{array}
    \right)^\flat
\end{array}
\end{logrel}
For $c_{\glu{\Pi}~A~B}$, fix a closed argument $M^\circ$ and its witness
$M^\bullet$. Reindex $c_B$ at
$(\spair{\gamma^\circ}{M^\circ},(\gamma^\bullet,M^\bullet))$
along $F\mapsto\syn{\kw{app}}~F~M^\circ$, then take dependent products
over both arguments using \maincref{lem:contrav-reindex,lem:contrav-pi}.

\begin{logrel}
\begin{array}{@{}l@{}}
  \glu{\kw{LAM}} :
    \glu{\kw{TM}}~(\Gamma \gext A)~B \to
    \glu{\kw{TM}}~\Gamma~(\glu{\Pi}~A~B)
  \\[3pt]
  (\glu{\kw{LAM}}~N)^\circ
    \isdef \syn{\kw{lam}}~N^\circ
  \\
  (\glu{\kw{LAM}}~N)^\bullet~
    \gamma^\circ~\gamma^\bullet~M^\circ~M^\bullet
    \isdef
    (\ssubst{\syn{\Pi_\beta}}
      {\spair{\gamma^\circ}{M^\circ}})^*
    \bigl(
      N^\bullet~
        \spair{\gamma^\circ}{M^\circ}~
        (\gamma^\bullet,M^\bullet)
    \bigr)
  \\[8pt]
  \glu{\kw{APP}} :
    (F : \glu{\kw{TM}}~\Gamma~(\glu{\Pi}~A~B)) \to
    (M : \glu{\kw{TM}}~\Gamma~A) \to
    \glu{\kw{TM}}~\Gamma~
      (B\glu{[}\glu{(}\glu{\kw{ID}}\glu{,}M\glu{)}\glu{]})
  \\[3pt]
  (\glu{\kw{APP}}~F~M)^\circ
    \isdef \syn{\kw{app}}~F^\circ~M^\circ
  \\
  (\glu{\kw{APP}}~F~M)^\bullet~
    \gamma^\circ~\gamma^\bullet
    \isdef
    F^\bullet~\gamma^\circ~\gamma^\bullet~
      (\ssubst{M^\circ}{\gamma^\circ})~
      (M^\bullet~\gamma^\circ~\gamma^\bullet)
\end{array}
\end{logrel}

\begin{samepage}
\subsubsection{Booleans}

The Boolean type definitions are:
\begin{logrel}
\begin{array}{@{}l@{}}
  \glu{\kw{BOOL}} :
    \glu{\kw{TY}}~\Gamma
  \\[3pt]
  \glu{\kw{BOOL}}^\circ
    \isdef \syn{\kw{Bool}}
  \\
  \glu{\kw{BOOL}}^\bullet_\flat~\gamma^\circ~\gamma^\bullet
    \isdef
    \bigl(\lambda M.\;
      \displaystyle
      \sum_{b : \{0,1\}}
        M \leq_{\stm~\syn{\emp}~\syn{\kw{Bool}}}
        \ulcorner b \urcorner
    \bigr)^\flat
  \\[3pt]
  \glu{\kw{BOOL}}^\bullet~\gamma^\circ~\gamma^\bullet~M
    \isdef
    \displaystyle
    \sum_{b : \{0,1\}}
      M \leq_{\stm~\syn{\emp}~\syn{\kw{Bool}}}
      \ulcorner b \urcorner
  \\[8pt]
  \glu{\kw{TRUE}}^\circ
    \isdef \syn{\kw{true}}
  \\[3pt]
  \glu{\kw{TRUE}}^\bullet~
    \gamma^\circ~\gamma^\bullet
    \isdef
    (1,\kw{id}_{\syn{\kw{true}}})
  \\[8pt]
  \glu{\kw{FALSE}}^\circ
    \isdef \syn{\kw{false}}
  \\[3pt]
  \glu{\kw{FALSE}}^\bullet~
    \gamma^\circ~\gamma^\bullet
    \isdef
    (0,\kw{id}_{\syn{\kw{false}}})
\end{array}
\end{logrel}
\end{samepage}
The proof of contravariance, $c_{\glu{\kw{BOOL}}}$, uses the representable families
$M\mapsto M\leq\ulcorner b\urcorner$
(\maincref{lem:representable-contravariant}) and discreteness of $\{0,1\}$. Its induced transport along a reduction $s : M \leq M'$ is
$s^*(b,r)=(b,s\cdot r)$.
For the dependent Boolean eliminator, with $C : \glu{\kw{TY}}~(\Gamma \gext \glu{\kw{BOOL}})$ and
$T : \glu{\kw{TM}}~\Gamma~\glu{\kw{BOOL}}$, if $T^\bullet~\gamma^\circ~\gamma^\bullet = (b,r)$, the eliminator is:
\begin{logrel}
\begin{array}{@{}l@{}}
  C_{\kw{true}}
    \isdef
    C\glu{[}\glu{(}\glu{\kw{ID}}\glu{,}
      \glu{\kw{TRUE}}\glu{)}\glu{]}
  \\
  C_{\kw{false}}
    \isdef
    C\glu{[}\glu{(}\glu{\kw{ID}}\glu{,}
      \glu{\kw{FALSE}}\glu{)}\glu{]}
  \\
  C_T
    \isdef
    C\glu{[}\glu{(}\glu{\kw{ID}}\glu{,}T\glu{)}\glu{]}
  \\[8pt]
  \glu{\kw{IF}} :
    (C : \glu{\kw{TY}}~
      (\Gamma \gext \glu{\kw{BOOL}})) \to
    \glu{\kw{TM}}~\Gamma~C_{\kw{true}} \to
    \glu{\kw{TM}}~\Gamma~C_{\kw{false}} \to
    (T : \glu{\kw{TM}}~\Gamma~\glu{\kw{BOOL}}) \to
    \glu{\kw{TM}}~\Gamma~C_T
  \\[3pt]
  (\glu{\kw{IF}}~C~U~V~T)^\circ
    \isdef
    \syn{\kw{if}}~C^\circ~U^\circ~V^\circ~T^\circ
  \\[8pt]
  (\glu{\kw{IF}}~C~U~V~T)^\bullet~\gamma^\circ~\gamma^\bullet
  \isdef
  \begin{cases}
    \rho_1^*\bigl(\kw{transport}(\kw{substInv}^{-1})
      (U^\bullet~\gamma^\circ~\gamma^\bullet)\bigr)
      & \text{if } b=1,\\
    \rho_0^*\bigl(\kw{transport}(\kw{substInv}^{-1})
      (V^\bullet~\gamma^\circ~\gamma^\bullet)\bigr)
      & \text{if } b=0.
  \end{cases}
\end{array}
\end{logrel}
The corresponding Boolean $\beta$-reduction gives, for $b = 1$ and $b = 0$ respectively,
\[
\begin{aligned}
  \rho_1 &: {}
  \ssubst{(\syn{\kw{if}}~C^\circ~U^\circ~V^\circ~T^\circ)}{\gamma^\circ}
  \leq
  \kw{conv}_{\ssubst{C^\circ}{\spair{\kw{id}_{\gamma^\circ}}{r}}}^{-1}~\ssubst{U^\circ}{\gamma^\circ},\\
  \rho_0 &: {}
  \ssubst{(\syn{\kw{if}}~C^\circ~U^\circ~V^\circ~T^\circ)}{\gamma^\circ}
  \leq
  \kw{conv}_{\ssubst{C^\circ}{\spair{\kw{id}_{\gamma^\circ}}{r}}}^{-1}~\ssubst{V^\circ}{\gamma^\circ}.
\end{aligned}
\]
In each branch, $\kw{substInv}$ comes from the comparison
\[
  \bigl(\spair{\gamma^\circ}{\ssubst{T^\circ}{\gamma^\circ}},
    (\gamma^\bullet,(b,r))\bigr)
  \leq
  \bigl(\spair{\gamma^\circ}{\ulcorner b\urcorner},
    (\gamma^\bullet,(b,\kw{id}_{\ulcorner b\urcorner}))\bigr)
\]
as in \cref{sec:universe:dependency-revisited}.

\subsubsection{Universe}

The universe type uses candidates, \ie contravariant flat predicates:
\begin{logrel}
\begin{array}{@{}l@{\;}c@{\;}l@{}}
  \Cand(A^\circ)
  &\isdef&
  \displaystyle
  \sum_{A^\bullet_\flat : \Pred(A^\circ)}
    \kw{isContrav}~(\eps(A^\bullet_\flat)).
\end{array}
\end{logrel}

\begin{samepage}
The universe definitions are:
\begin{logrel}
\begin{array}{@{}l@{}}
  \glu{\kw{UNIV}}^\circ
    \isdef \syn{\kw{U}},
  \\[3pt]
  \glu{\kw{UNIV}}^\bullet_\flat~\gamma^\circ~\gamma^\bullet
    \isdef
    \bigl(\lambda M.\; \Cand(\syn{\kw{El}}~M)\bigr)^\flat
  \\[8pt]
  (\glu{\kw{EL}}~M)^\circ
    \isdef \syn{\kw{El}}~M^\circ,
  \\[3pt]
  (\glu{\kw{EL}}~M)^\bullet_\flat~\gamma^\circ~\gamma^\bullet
    \isdef \pi_1(M^\bullet~\gamma^\circ~\gamma^\bullet),
  \\[3pt]
  c_{\glu{\kw{EL}}~M}~\gamma^\circ~\gamma^\bullet
    \isdef \pi_2(M^\bullet~\gamma^\circ~\gamma^\bullet)
  \\[8pt]
  (\glu{\kw{CODE}}~A)^\circ
    \isdef
    \syn{\kw{Code}}~A^\circ,
  \\[3pt]
  (\glu{\kw{CODE}}~A)^\bullet~
    \gamma^\circ~\gamma^\bullet
  \isdef
  (\ssubst{\syn{\kw{U}_{\beta}}}{\gamma^\circ})^*
    \bigl(A^\bullet_\flat~\gamma^\circ~\gamma^\bullet,\,
          c_A~\gamma^\circ~\gamma^\bullet\bigr)
\end{array}
\end{logrel}
\end{samepage}
The fibers of $\Cand$ are discrete, since $\Pred(A^\circ)$ is flat and
$\kw{isContrav}$ is a proposition. By \maincref{lem:contrav-discrete-base},
$\Cand$ is contravariant over the discrete base $\sty~\syn{\emp}$;
reindexing along $\syn{\kw{El}}$ gives $c_{\glu{\kw{UNIV}}}$
(\maincref{lem:contrav-reindex}).

\subsection{Binary Model}

This part records the binary structures used in \maincref{sec:binary}.
For binary families, $\kw{isContrav}$ refers to the uncurried family.

\subsubsection{Binary Predicates}

For closed syntactic types
$A_L^\circ,A_R^\circ : \sty~\syn{\emp}$, the binary predicates are:
\begin{logrel}
\begin{array}{@{}l@{}}
  \Pred_2(A_L^\circ,A_R^\circ)
  \isdef
  \kw{let}^{\flat}~B_L = \kappa(A_L^\circ)~\kw{in}~
  \kw{let}^{\flat}~B_R = \kappa(A_R^\circ)~\kw{in}~
  \flat(\stm~\syn{\emp}~B_L \to \stm~\syn{\emp}~B_R \to \mathcal{V}).
\end{array}
\end{logrel}

\subsubsection{Binary Judgmental Records}

\begin{logrel}
\begin{array}{@{\guiderule{oc-blue-9}}l@{}}
  \declkw{record}~\glu{\kw{CTX}_2} : \mathcal{U}_s~\declkw{where} \\
  \quad \Gamma_L^\circ,\,\Gamma_R^\circ : \sctx \\
  \quad \Gamma^\bullet_\flat :
    \flat(\ssub~\syn{\emp}~\Gamma_L^\circ
      \to \ssub~\syn{\emp}~\Gamma_R^\circ \to \mathcal{V}) \\
  \quad \Gamma^\bullet \isdef \eps(\Gamma^\bullet_\flat) \\
  \quad c_\Gamma : \kw{isContrav}~\Gamma^\bullet
  \\ \noalign{\vskip 8pt}
  \declkw{record}~\glu{\kw{SUB}_2}~
    (\Gamma~\Delta : \glu{\kw{CTX}_2}) : \mathcal{U}_p~\declkw{where} \\
  \quad \sigma_L^\circ : \ssub~\Gamma_L^\circ~\Delta_L^\circ \qquad
        \sigma_R^\circ : \ssub~\Gamma_R^\circ~\Delta_R^\circ \\
  \quad \sigma^\bullet :
    (\gamma_L^\circ : \ssub~\syn{\emp}~\Gamma_L^\circ)
    \to (\gamma_R^\circ : \ssub~\syn{\emp}~\Gamma_R^\circ)
    \to \Gamma^\bullet~\gamma_L^\circ~\gamma_R^\circ \\
  \qquad \to \Delta^\bullet~
    (\sigma_L^\circ\scomp\gamma_L^\circ)~(\sigma_R^\circ\scomp\gamma_R^\circ)
  \\ \noalign{\vskip 8pt}
  \declkw{record}~\glu{\kw{TY}_2}~
    (\Gamma : \glu{\kw{CTX}_2}) : \mathcal{U}_s~\declkw{where} \\
  \quad A_L^\circ : \sty~\Gamma_L^\circ \qquad A_R^\circ : \sty~\Gamma_R^\circ \\
  \quad A^\bullet_\flat :
    (\gamma_L^\circ : \ssub~\syn{\emp}~\Gamma_L^\circ)
    \to (\gamma_R^\circ : \ssub~\syn{\emp}~\Gamma_R^\circ)
    \to \Gamma^\bullet~\gamma_L^\circ~\gamma_R^\circ \\
  \qquad \to \Pred_2(\ssubst{A_L^\circ}{\gamma_L^\circ},
      \ssubst{A_R^\circ}{\gamma_R^\circ}) \\
  \quad A^\bullet~\gamma_L^\circ~\gamma_R^\circ~\gamma^\bullet
    \isdef
    \eps(A^\bullet_\flat~
      \gamma_L^\circ~\gamma_R^\circ~\gamma^\bullet) \\
  \quad c_A :
    \forall\,\gamma_L^\circ~\gamma_R^\circ~\gamma^\bullet.\;
    \kw{isContrav}~
      (A^\bullet~\gamma_L^\circ~\gamma_R^\circ~\gamma^\bullet)
  \\ \noalign{\vskip 8pt}
  \declkw{record}~\glu{\kw{TM}_2}~
    (\Gamma : \glu{\kw{CTX}_2})~
    (A : \glu{\kw{TY}_2}~\Gamma) : \mathcal{U}_p~\declkw{where} \\
  \quad M_L^\circ : \stm~\Gamma_L^\circ~A_L^\circ \qquad
        M_R^\circ : \stm~\Gamma_R^\circ~A_R^\circ \\
  \quad M^\bullet :
    (\gamma_L^\circ : \ssub~\syn{\emp}~\Gamma_L^\circ)
    \to (\gamma_R^\circ : \ssub~\syn{\emp}~\Gamma_R^\circ)
    \to (\gamma^\bullet :
      \Gamma^\bullet~\gamma_L^\circ~\gamma_R^\circ) \\
  \qquad \to A^\bullet~\gamma_L^\circ~\gamma_R^\circ~\gamma^\bullet~
      (\ssubst{M_L^\circ}{\gamma_L^\circ})~
      (\ssubst{M_R^\circ}{\gamma_R^\circ})
\end{array}
\end{logrel}

\subsubsection{Binary Type-Former Examples}

The binary Boolean predicate records a common Boolean value:
\begin{logrel}
\begin{array}{@{}l@{}}
  \glu{\kw{BOOL}_2} :
    \glu{\kw{TY}_2}~\Gamma
  \\
  (\glu{\kw{BOOL}_2})_i^\circ
    \isdef \syn{\kw{Bool}} \qquad (i\in\{L,R\})
  \\
  \glu{\kw{BOOL}_2}^\bullet_\flat~
    \gamma_L^\circ~\gamma_R^\circ~\gamma^\bullet
    \isdef
    \left(
      \begin{array}{@{}l@{}}
        \lambda M_L~M_R.\;
        \displaystyle
        \sum_{b : \{0,1\}}
          \bigl(
            M_L \leq_{\stm~\syn{\emp}~\syn{\kw{Bool}}}
              \ulcorner b \urcorner
          \bigr) \times
          \bigl(
            M_R \leq_{\stm~\syn{\emp}~\syn{\kw{Bool}}}
              \ulcorner b \urcorner
          \bigr)
      \end{array}
    \right)^\flat
\end{array}
\end{logrel}

For $c_{\glu{\kw{BOOL}_2}}$, use products of representable families
indexed by the discrete type $\{0,1\}$, as in the unary case. The induced
transport along reductions $s_L$ and $s_R$ is
$(s_L,s_R)^*(b,(r_L,r_R))=(b,(s_L\cdot r_L,s_R\cdot r_R))$.

\begin{samepage}
The binary product predicate is pointwise on projections:
\begin{logrel}
\begin{array}{@{}l@{}}
  \glu{\kw{PROD}_2} :
    \glu{\kw{TY}_2}~\Gamma \to
    \glu{\kw{TY}_2}~\Gamma \to
    \glu{\kw{TY}_2}~\Gamma
  \\
  (\glu{\kw{PROD}_2}~A~B)_i^\circ
    \isdef A_i^\circ \mathbin{\syn{\times}} B_i^\circ \qquad (i\in\{L,R\})
  \\
  (\glu{\kw{PROD}_2}~A~B)^\bullet_\flat~
    \gamma_L^\circ~\gamma_R^\circ~\gamma^\bullet
    \isdef
    \kw{let}^{\flat}~A^\bullet =
      A^\bullet_\flat~
        \gamma_L^\circ~\gamma_R^\circ~\gamma^\bullet~\kw{in}~
    \kw{let}^{\flat}~B^\bullet =
      B^\bullet_\flat~
        \gamma_L^\circ~\gamma_R^\circ~\gamma^\bullet~\kw{in}~
    \\
    \quad
    \left(
      \begin{array}{@{}l@{}}
        \lambda P_L~P_R.\;
          A^\bullet~(\syn{\kw{fst}}~P_L)~
            (\syn{\kw{fst}}~P_R)
          \times
          B^\bullet~(\syn{\kw{snd}}~P_L)~
            (\syn{\kw{snd}}~P_R)
      \end{array}
    \right)^\flat
\end{array}
\end{logrel}
\end{samepage}

\subsubsection{Binary Universe}

The binary universe uses binary candidates:
\begin{logrel}
\begin{array}{@{}l@{}}
  \Cand_2(A_L^\circ,A_R^\circ)
  \isdef
  \displaystyle
  \sum_{A^\bullet_\flat : \Pred_2(A_L^\circ,A_R^\circ)}
    \kw{isContrav}~(\eps(A^\bullet_\flat)).
\end{array}
\end{logrel}

The binary universe is:
\begin{logrel}
\begin{array}{@{}l@{}}
  (\glu{\kw{UNIV}_2})_i^\circ \isdef \syn{\kw{U}} \qquad (i\in\{L,R\})
  \\
  \glu{\kw{UNIV}_2}^\bullet_\flat~
    \gamma_L^\circ~\gamma_R^\circ~\gamma^\bullet
  \isdef
  \bigl(\lambda M_L~M_R.\;
    \Cand_2(\syn{\kw{El}}~M_L,\syn{\kw{El}}~M_R)\bigr)^\flat
\end{array}
\end{logrel}

As in the unary case, $\Cand_2$ has discrete fibers over the discrete base
$(\sty~\syn{\emp})\times(\sty~\syn{\emp})$. Thus
\maincref{lem:contrav-discrete-base,lem:contrav-reindex} give
$c_{\glu{\kw{UNIV}_2}}$ by reindexing along
$(M_L,M_R)\mapsto(\syn{\kw{El}}~M_L,\syn{\kw{El}}~M_R)$.
 }{}

\ifappendixonly\paperreferences\fi
\end{document}